\documentclass[12pt,a4paper]{article}
\usepackage[german, english]{babel}
\usepackage{a4wide}
\usepackage{amsmath}
\usepackage{amssymb}
\usepackage{graphicx}
\usepackage{ifthen}
\usepackage{epsfig}
\usepackage{setspace}
\newcounter{fig}   \newcommand{\lbfig}[1]{\refstepcounter{fig}
\label{#1} }

\newcommand{\bea}{\begin{eqnarray}}
\newcommand{\eea}{\end{eqnarray}}
\newcommand{\be}{\begin{equation}}
\newcommand{\ee}{\end{equation}}

\def\bfph{{\pmb{\phi}}}

\newcommand{\re}[1]{(\ref{#1})}

\begin{document}
\title{Isorotating Baby Skyrmions}
\author{A. Halavanau$^{\dagger}$ and Ya. Shnir$^{\dagger \ddagger \star}$\\[10pt]
\small{$^{\dagger}$Department of Theoretical Physics and Astrophysics,
Belarusian State University, Minsk,  Belarus}\\
\small{$^{\ddagger}$BLTP, Dubna, Russia}\\
\small{$^{\star}$Institute of Physics, Carl von Ossietzky University Oldenburg, Germany}}

%\author{A. Halavanau$^{\dagger}$ and Ya. Shnir$^{\dagger \star}$}
%\affiliation{$^{\dagger}$Department of Theoretical Physics and Astrophysics,
%Belarusian State University, Minsk,  Belarus\\
%$^{\star}$Institute of Physics, Carl von Ossietzky University Oldenburg, Germany}

\date{\today}
%\pacs{04.20.Jb, 04.40.Nr}

\maketitle

\begin{abstract}
{\sf We discuss how internal rotation with fixed angular frequency can affect the solitons in the
baby Skyrme model in which the global $O(3)$ symmetry is broken to the $SO(2)$. Two particular
choices of the potential term are considered, the "old" potential and the
"new" double vacuum potential, We do not impose any
assumptions about the symmetry on the fields.
Our results confirm existence of two types of instabilities determined by the
relation between the mass parameter of the potential $\mu$ and the angular frequency $\omega$.
It is shown that multiSkyrmions in the model with "old" potential at some critical value of the angular frequency
become unstable with respect to decay into single
Skyrmion constituents.}
\end{abstract}
%\medskip

%%%%%%%%%%%%%%%%%%%%%%%%%%%%%%%%%%%%%%%%%%%%%%%%%%%%%%%%%%%%%%%%%%
\section{Introduction}
%%%%%%%%%%%%%%%%%%%%%%%%%%%%%%%%%%%%%%%%%%%%%%%%%%%%%%%%%%%%%%%%%%

The so-called baby Skyrme model is a modified version of the non-linear $O(3)$ $\sigma$-model in $2+1$ dimensions \cite{Bsk},
a low-dimensional simplified theory which resembles the conventional Skyrme model in many important respects.
This model has a number of applications, e.g. in condensed matter physics where Skyrmion configurations were
observed experimentally
\cite{CondMatt}, or in the topological quantum Hall effect \cite{SkHall}.

Together with the original Skyrme model in $d=3+1$ \cite{Skyrme}  and the Faddeev--Skyrme model \cite{Faddeev},
the baby Skyrme model can be considered as a member of the Skyrme family. Indeed, the Lagrangian of all these models
has similar structure, it includes the usual $O(3)$ sigma model kinetic term,
the Skyrme term, which is quartic in derivatives, and the potential term which does not
contain the derivatives.

According to the Derrick's theorem \cite{Derrick} the Skyrme term is always required to support
existence of soliton configurations. In $d=2+1$ the potential term is also necessary to obtain the
localised field configurations with finite energy.
In $d=3+1$ the last term is optional, it gives a mass to the excitations of the scalar field, so
in the context of the usual Skyrme model it is referred to as "pion mass term".
It is known that it appearance might seriously affect the
structure of the solutions of the Skyrme model \cite{Battye}.
On the other hand, the form of the potential term in the baby Skyrme model is largely arbitrary, there are different choices
related with various ways of symmetry breaking \cite{Ward,Hen,JSS}.

Note that the solitons of the models of the Skyrme family possess both rotational and
internal rotational (or isorotational) degrees of freedom.
Traditional approach to study the spinning solitons is related with rigid body approximation, both in the context of the Skyrme model
\cite{ANW,Adkins:1983hy} and in the baby Skyrme model \cite{PZS}. The assumption is that the solitons could
rotate without changing its shape. This restriction can be weakly relaxed by consideration of
the radial deformations which would not violated the rotational symmetry of the hedgehog configuration \cite{PZS,Braaten}. Evidently,
this approximation is not very satisfactory, a consistent approach is to solve full system of field equation without imposing any
spatial symmetries on the isospinning solitons. Furthermore, almost all previous studies of spinning solitons
(see e.g. \cite{BattyMareike,Battye:2005nx}) were concerned with minimization of the total energy functional $E_J[\bfph]$ for fixed
value of the isospin $J$. However if we do not assume the spinning soliton will have precisely the same shape
as the static soliton, this approach becomes rather involved, it
is related with numerical solution of complicated differential-integral equation.

Very recently the isospinning  soliton solutions were considered in the Faddeev–Skyrme model beyond rigid body
approximation \cite{BattyMareike,JHSS}.
The approach of the paper \cite{JHSS} is to consider the static pseudo-energy minimization problem, where
the pseudo-energy functional  $F_\omega[\bfph]$ depends parametrically on the angular frequency $\omega$.
The important conclusion which
is general for all models of the Skyrme family, is that there is a new type of instability of the solitons
due to the extra nonlinear velocity dependence generated by the Skyrme term \cite{JHSS}.

In this letter, we aim to perform the analysis of the critical behavior of the isospinning solitons of the baby
Skyrme model without imposing any spacial symmetries. We confirm existence of two types of instabilities determined by the
relation between the mass parameter of the potential $\mu$ and the frequency $\omega$ similar to the pattern observed in the
Faddeev--Skyrme model \cite{BattyMareike,JHSS}. Interestingly, we observe that the critical behavior of the isospinning baby Skyrmions
depends also on the structure of the potential of the model, for example in the case of the "old" model \cite{Bsk}
the isospinning configurations of higher degree may become unstable with respect to
decay into constituents.

Note that we do not consider very interesting case of the family of the potentials discussed by Karliner and Hen \cite{Hen}.
Here we also restrict our consideration to configurations of degree $1\le B \le 5$.  More systematic investigation of
the spinning baby Skyrmions for various potentials and for larger values of $B$ will be presented elsewhere in our work in
collaboration with R.~Battye and M.~Haberichter \cite{BHHS}.

In the next section we briefly
review our approach to the minimization problem of finding isospinning solitons
of the baby Skyrme model.
The numerical results are presented in
Section 3 where we considered behavior of the isorotating solitons in two different
cases, the  "old" baby Skyrme model \cite{Bsk}  and the "new" double vacuum model \cite{Weidig:1998ii}.
We give our conclusions and remarks in the final section.

At last but not least, we consider our results to be complementary to the independent findings of R.~Battye and M.~Haberichter
presented in the recent paper \cite{BattyMareike2}.

\section{Baby Skyrme model}
As a starting point we consider the rescaled Lagrangian\footnote{Our conventions are slightly different from usual choice \cite{Bsk},
the kinetic term differs from the standard one by a factor of two. Evidently, corresponding rescaling of the mass parameter
$\mu$ allows us to recover the latter conventions.}
of the $O(3)$ $\sigma$-model with the Skyrme term in $2+1$ dimensions \cite{Bsk}
\be \label{Lag}
L= \partial_\mu \bfph \cdot \partial^\mu \bfph - \frac{1}{4}(\partial_\mu \bfph
\times \partial_\nu
\bfph)^2 - U[\bfph]
\ee
where $\bfph = (\phi_1, \phi_2,\phi_3)$ denotes a triplet of scalar
fields which satisfy the constraint $|\bfph|^2=1$.
Topologically the field is the map
$\phi:\mathbb{R}^2 \to S^2$ characterized by the topological charge $B = \pi_2(S^2) = \mathbb{Z}$. Explicitly,
\be \label{charge}
B=\frac{1}{4\pi}\int \bfph \cdot \partial_1 \bfph \times \partial_2 \bfph ~d^2x
\ee

Note that the first two terms in the functional \re{Lag} are invariant under the global $O(3)$ transformations, this symmetry
becomes broken via the potential term. The standard choice of the potential of the baby Skyrme model is \cite{Bsk}
\be \label{pot-old}
U[\bfph] = \mu^2 [1-\phi_3]\, ,
\ee
thus the symmetry is broken to $SO(2)$ and there is a unique vacuum $\bfph_\infty = (0,0,1)$.  The corresponding solitons of degree
$B=1,2$ are axially symmetric \cite{Bsk} however the rotational symmetry of the configurations of higher degree
becomes broken \cite{PZS}.

The residual symmetry of the configurations with respect
to the rotations around the third axis in the internal space
allows us to consider the stationary isospinning (i.e. internally rotating) solitons
\be \label{rot}
(\phi_1 + i\phi_2) \mapsto (\phi_1 + i\phi_2)e^{i\omega t},\ ,
\ee
where $\omega$ is the angular frequency. The corresponding conserved quantity is the angular momentum $J = \omega \Lambda[\bfph]$,
where  $\Lambda[\bfph]$ is
the moment of inertia, thus the total energy of the spinning field configuration is
$$E_J[\bfph] = V[\bfph] + \frac{J^2}{2\Lambda[\bfph]} .
$$

Evidently, the isorotations \re{rot} of the energy functional of the baby Skyrme model yield the pseudo-energy functional
\be
\begin{split}
\label{F}
F_\omega[\bfph] &= \int\limits_{\mathbb{R}_2}\biggl\{\frac{1}{2}\left( [2-\omega^2(\bfph_\infty \times \bfph)^2 ]
(\partial_i \bfph \cdot \partial_i \bfph) + \omega^2 [{\bfph}_\infty \cdot (\bfph \times \partial_i \bfph) ]^2\right)\\
&+ \frac{1}{4}(\partial_i \bfph
\times \partial_j
\bfph)^2
+ \left( U[\bfph] - {\omega^2} ({\bfph}_\infty \times \bfph)^2 \right)
\biggr\}\\
&= V - \frac{1}{2}\omega^2 \Lambda(\bfph)
\end{split}
\ee
where the $V$ is the potential energy of the non-rotated configuration and the moment of inertia is
\be
\Lambda(\bfph) = \int\limits_{\mathbb{R}_2}\biggl\{ ({\bfph}_\infty \times \bfph)^2 [1+
(\partial_i \bfph \cdot \partial_i \bfph)] - [{\bfph}_\infty \cdot (\bfph \times {\partial_i} {\bfph}) ]^2\biggr\}
\ee
The isospinning solitons correspond to the stationary points of the functional \re{F}, it has the same dimensional structure
as the total energy functional, thus the standard scaling arguments in two spacial dimensions \cite{Derrick}
yield the virial relation
\be  \label{virial}
\mathbb{G}=\mathbb{V}
\ee
where $\mathbb{G} = \frac{1}{4} \int\limits_{\mathbb{R}_2} (\partial_i \bfph
\times \partial_j \bfph)^2 $ and $\mathbb{V} = \int\limits_{\mathbb{R}_2}
\left( U[\bfph] - {\omega^2} ({\bfph}_\infty \times \bfph)^2 \right)$ are two integrals which must be positively defined.

However the pseudoenergy \re{F} is not bounded from below for $\omega > \omega_1 = \sqrt 2$ independently from
the particular choice
of the potential $U[\bfph]$ \cite{JHSS}. Indeed, the first term in \re{F} effectively defines
the geometry of the deformed
sphere $S^2$ squashed along the direction $\bfph_\infty$, the metric on this space becomes
singular at $\omega=\omega_1=\sqrt 2$.

\begin{figure}
\lbfig{f-2}
\begin{center}
{\includegraphics[scale=0.25,angle=-90]{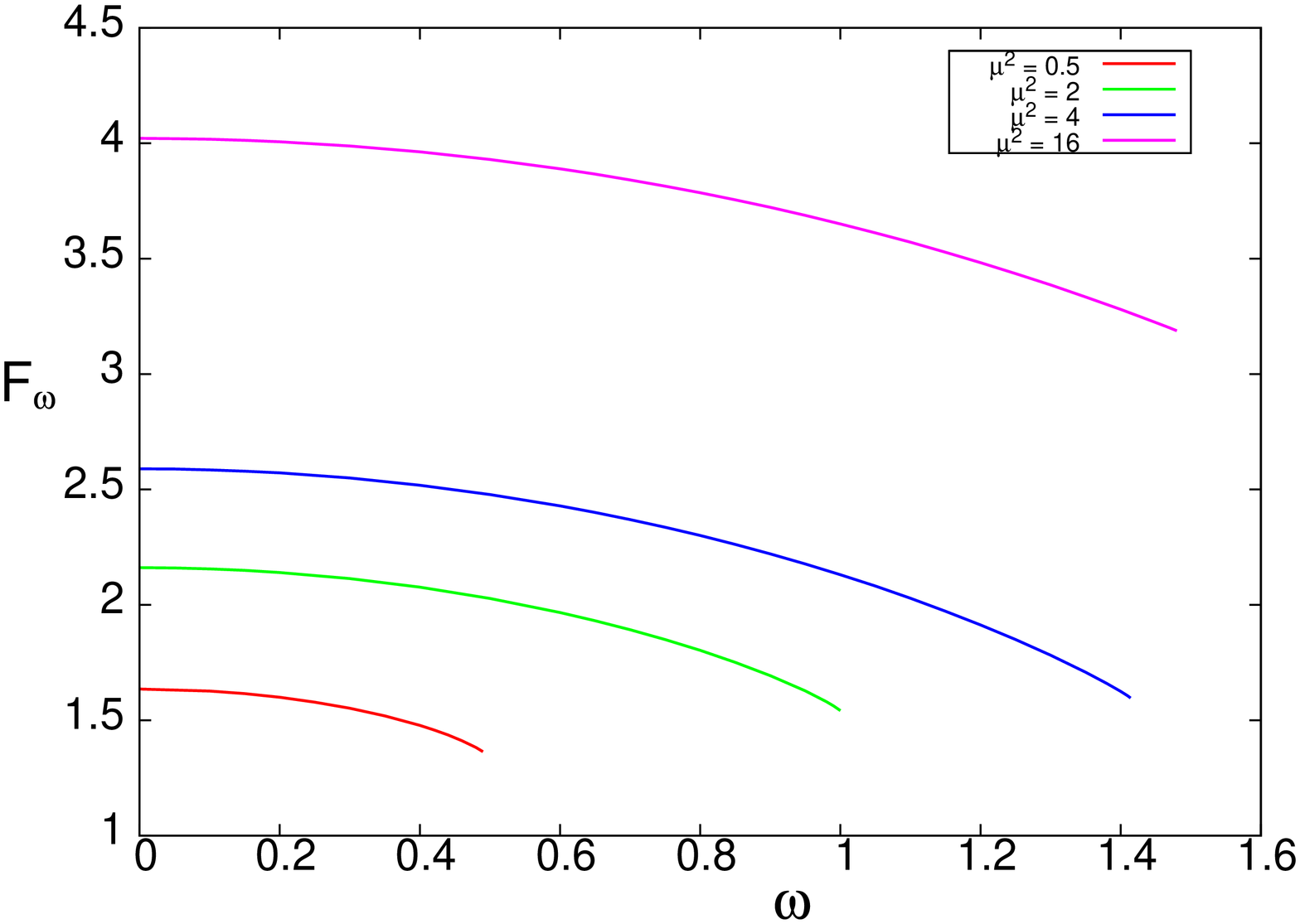}}
{\includegraphics[scale=0.25,angle=-90]{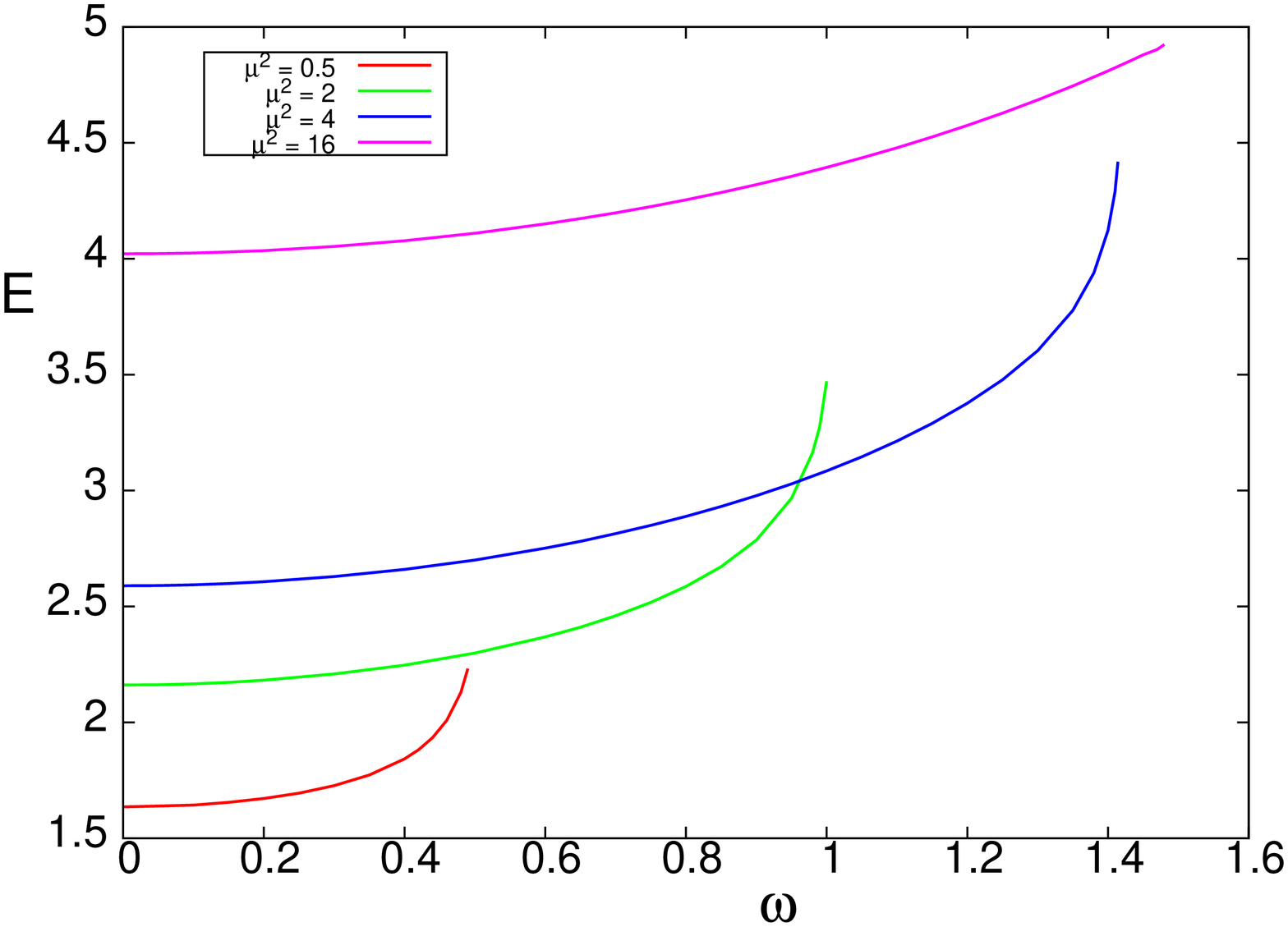}}
{\includegraphics[scale=0.25,angle=-90]{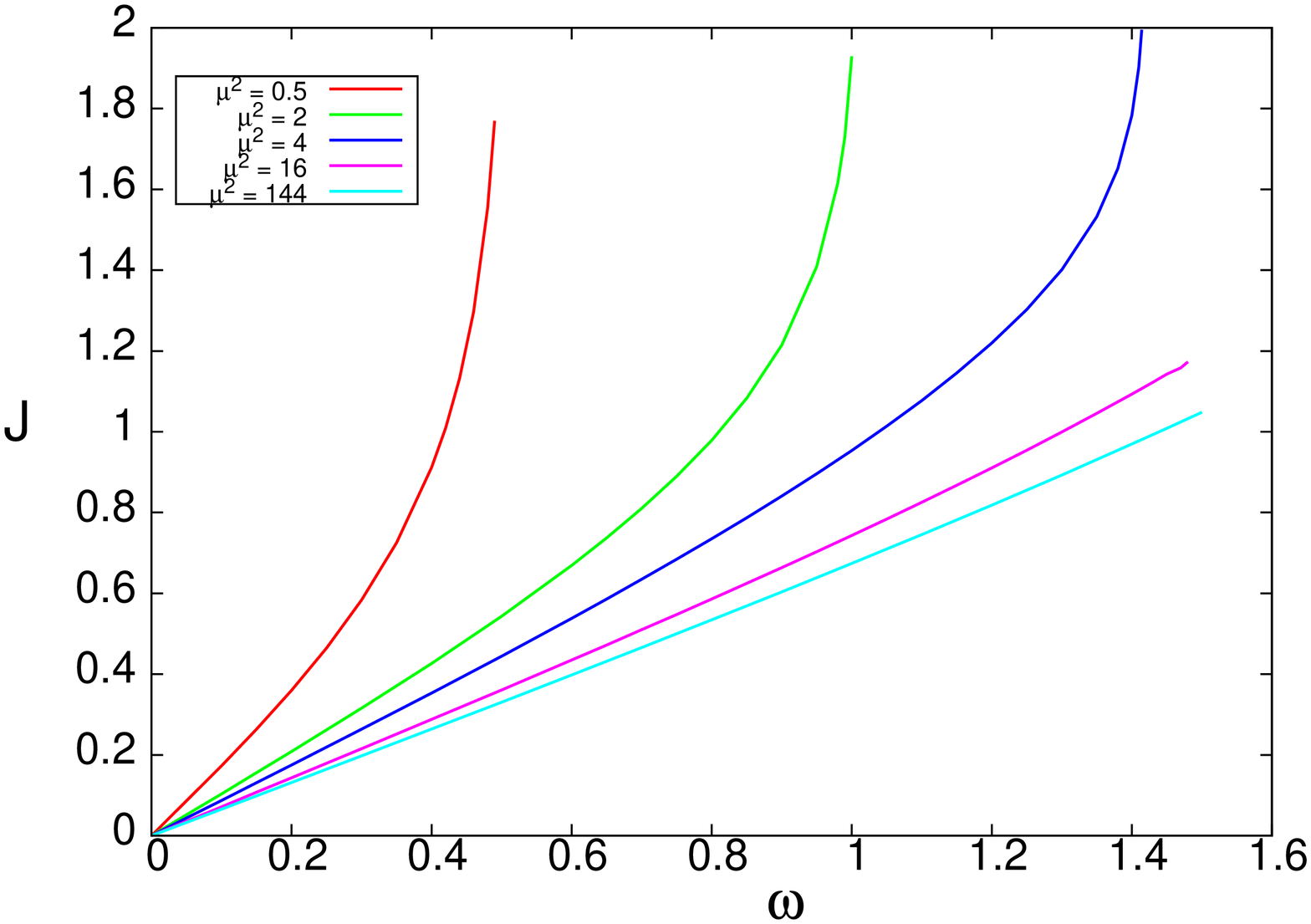}}
{\includegraphics[scale=0.25,angle=-90]{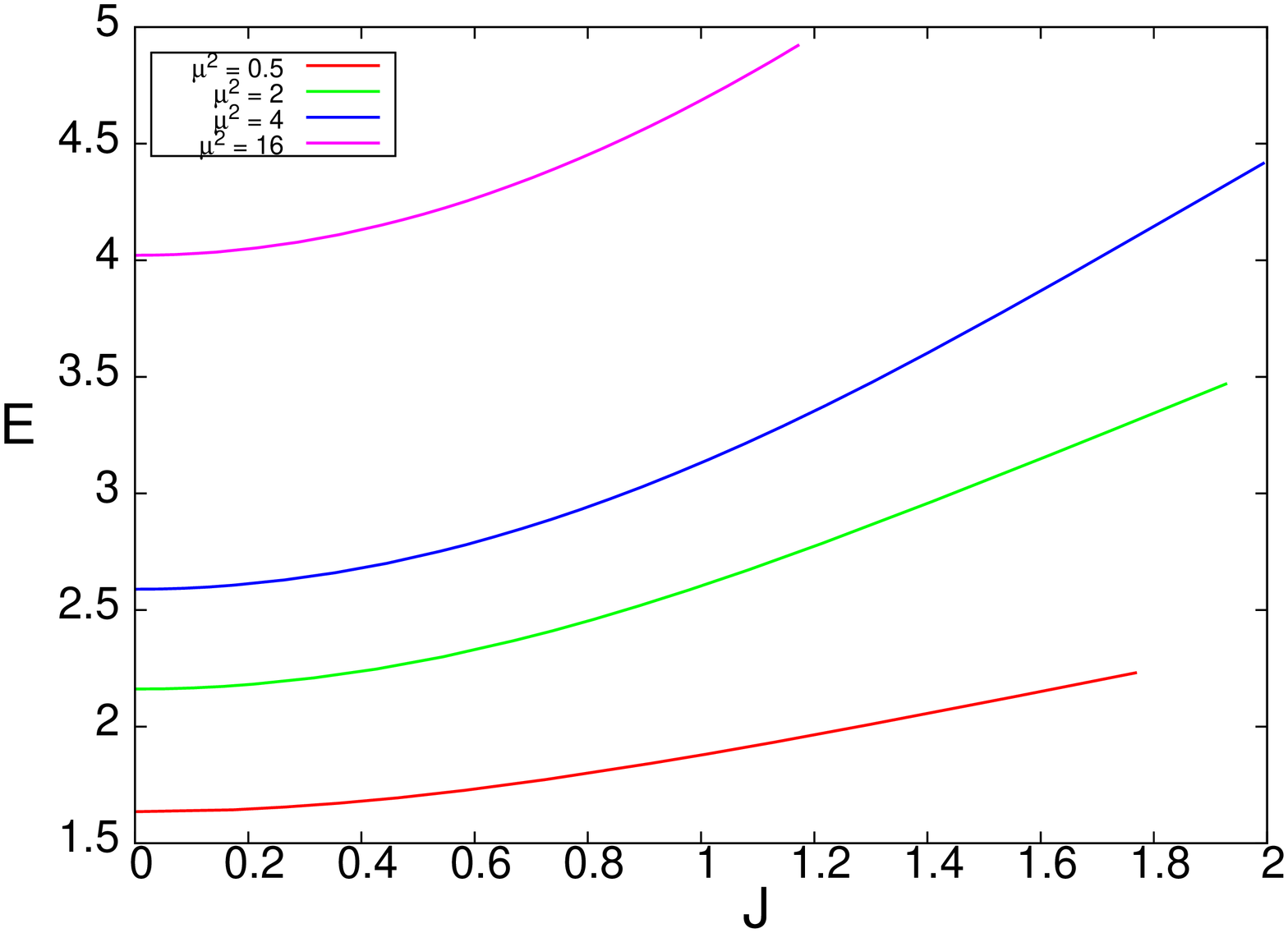}}
\end{center}
\caption{(Color online) Pseudoenergy, energy and isospin of the $B=1$ baby Skyrmion in the model with potential \re{pot-old}
are plotted as functions of angular frequency $\omega$, and the energy
is plotted as a function of isospin $J$ at $\mu^2=0.5,2,4,16$.}
\end{figure}

The second critical frequency is related with condition of positiveness of the effective
potential
$$
U_\omega[\bfph]= U[\bfph] - \omega^2(1-\phi_3^2) \, ,
$$ it approaches zero at some critical value $\omega = \omega_2$.
In this limit the isospinning solitons of the baby Skyrme model cease to exist because the vanishing of
the potential makes the configuration unstable, the virial relation \re{virial} becomes violated.
It is particularly convenient to investigate
the critical behavior of that type not
in the case of the potential \re{pot-old} but for the double vacuum model \cite{Weidig:1998ii} with
another choice of the rotationally invariant potential
\be
\label{double}
U[\bfph] = \mu^2 [1-\phi_3^2]\, ,
\ee
Evidently in that case the critical value $\omega_2 = \mu$. Below we consider pattern of critical behavior in both models.

The traditional approach to study the solitons of the model \re{Lag}
is related with
separation of the  radial and angular variables \cite{Bsk,PZS}, thus the consideration becomes restricted to the
case of rotationally invariant configurations and the corresponding Euler-Lagrange equations are reduced to a single
ordinary differential equation on radial function $f(\rho)$. However more detailed analyse reveal that the higher charge
 $B\ge 3$ baby Skyrmions may not possess rotational symmetry \cite{Bsk,Foster},  starting from some critical value of the
mass parameter $\mu$  the global minimum of the energy functional corresponds to the configurations with discrete
symmetries.

\begin{figure}
\lbfig{f-3}
\begin{center}
{\includegraphics[scale=0.24,angle=-90]{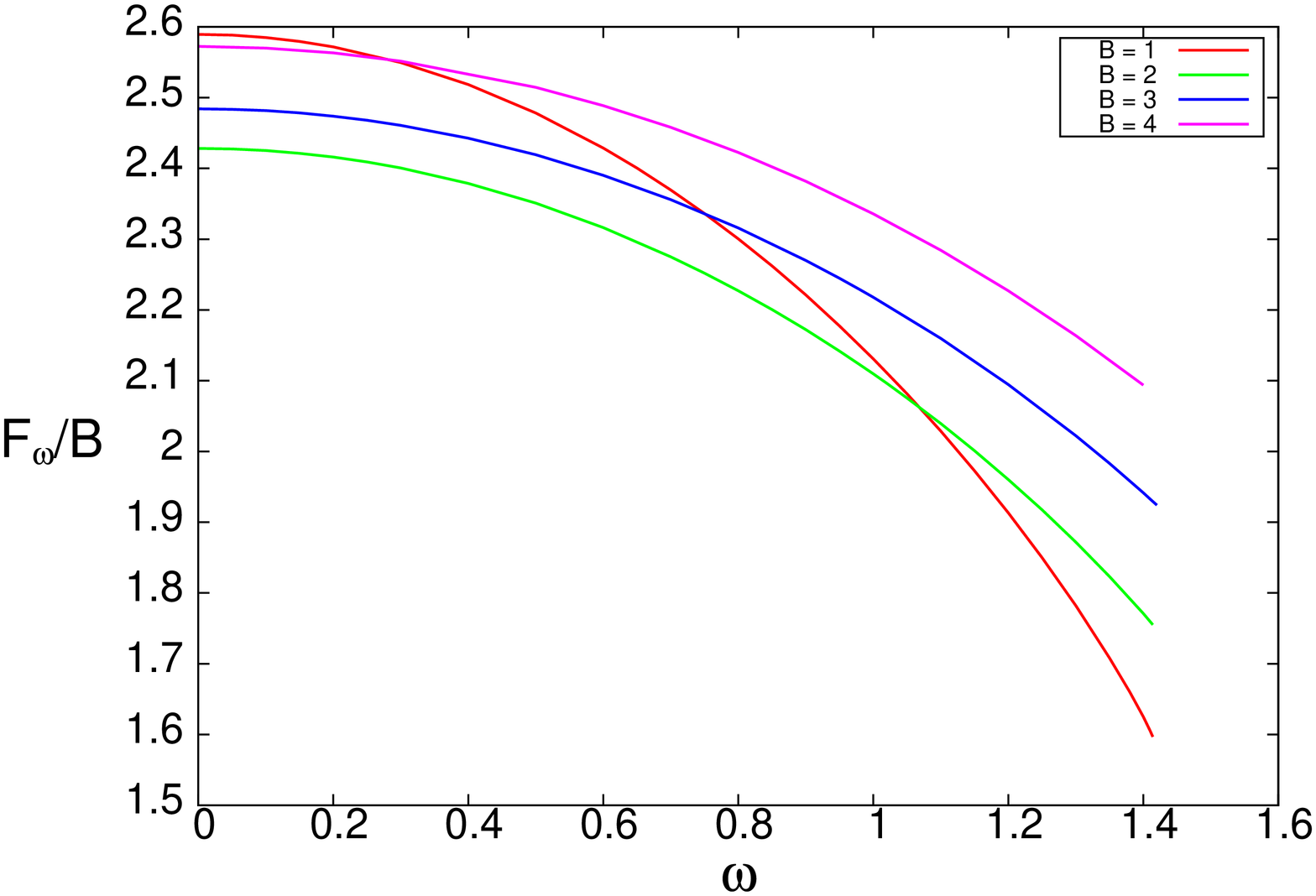}}
{\includegraphics[scale=0.24,angle=-90]{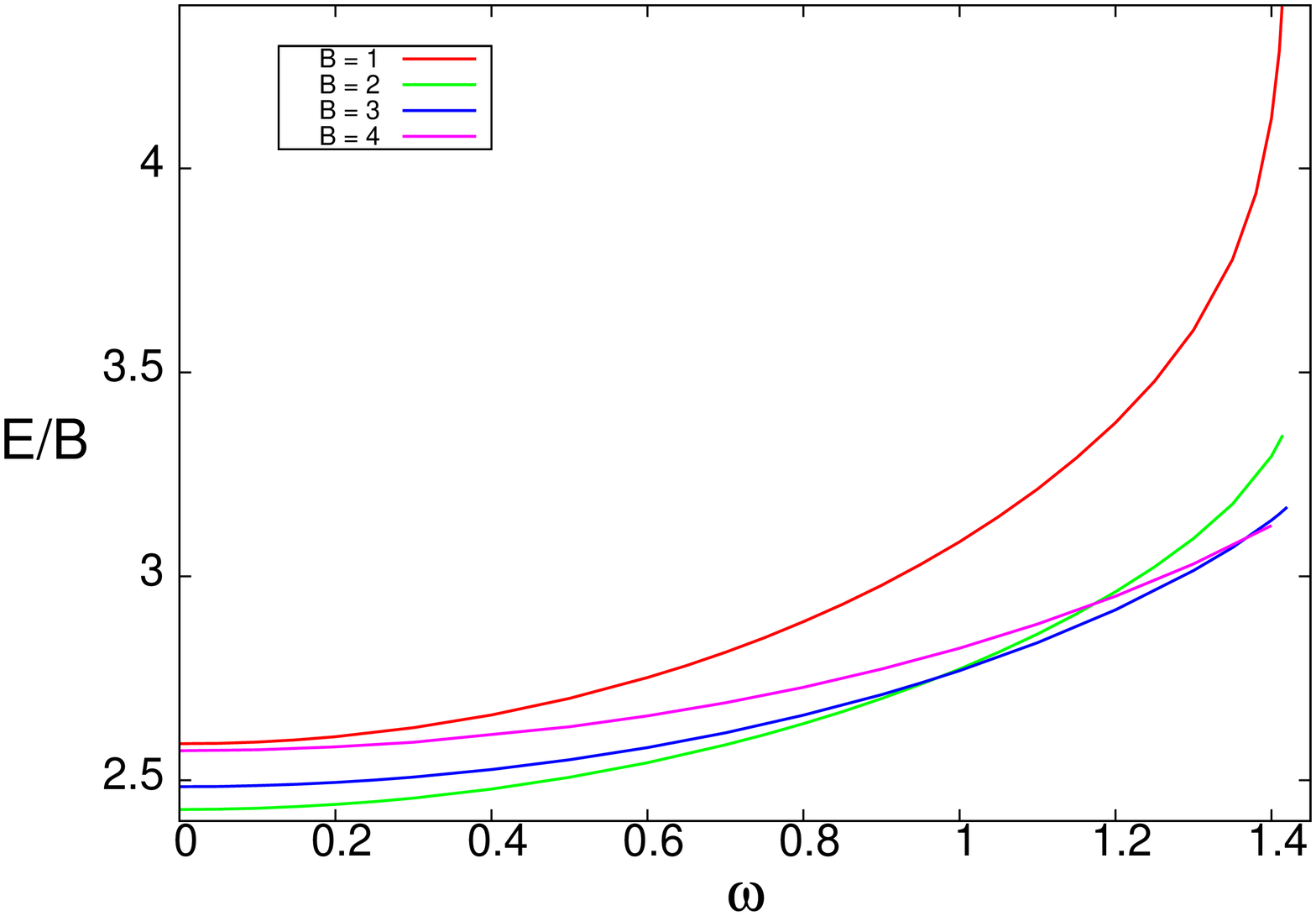}}
{\includegraphics[scale=0.24,angle=-90]{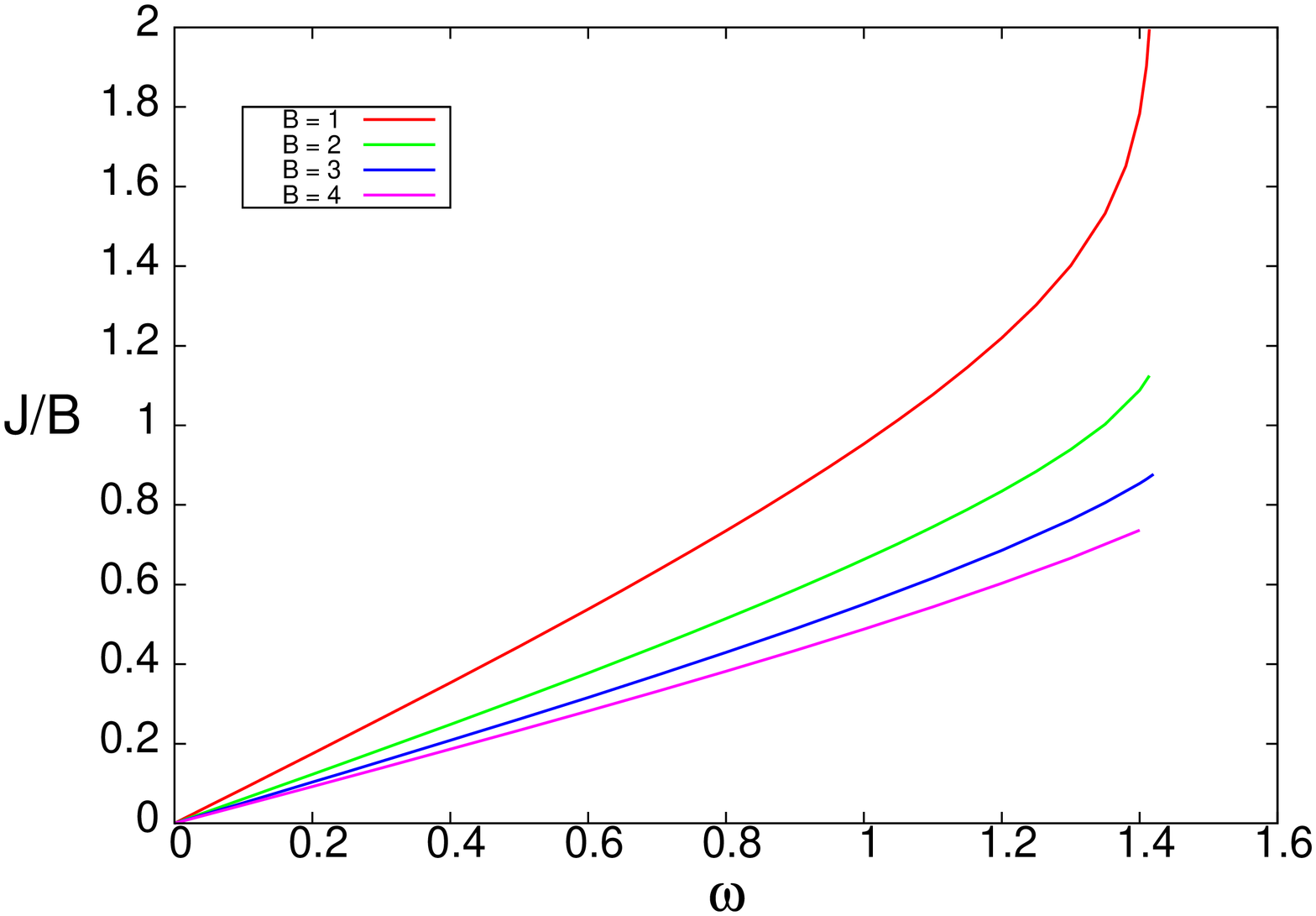}}
{\includegraphics[scale=0.24,angle=-90]{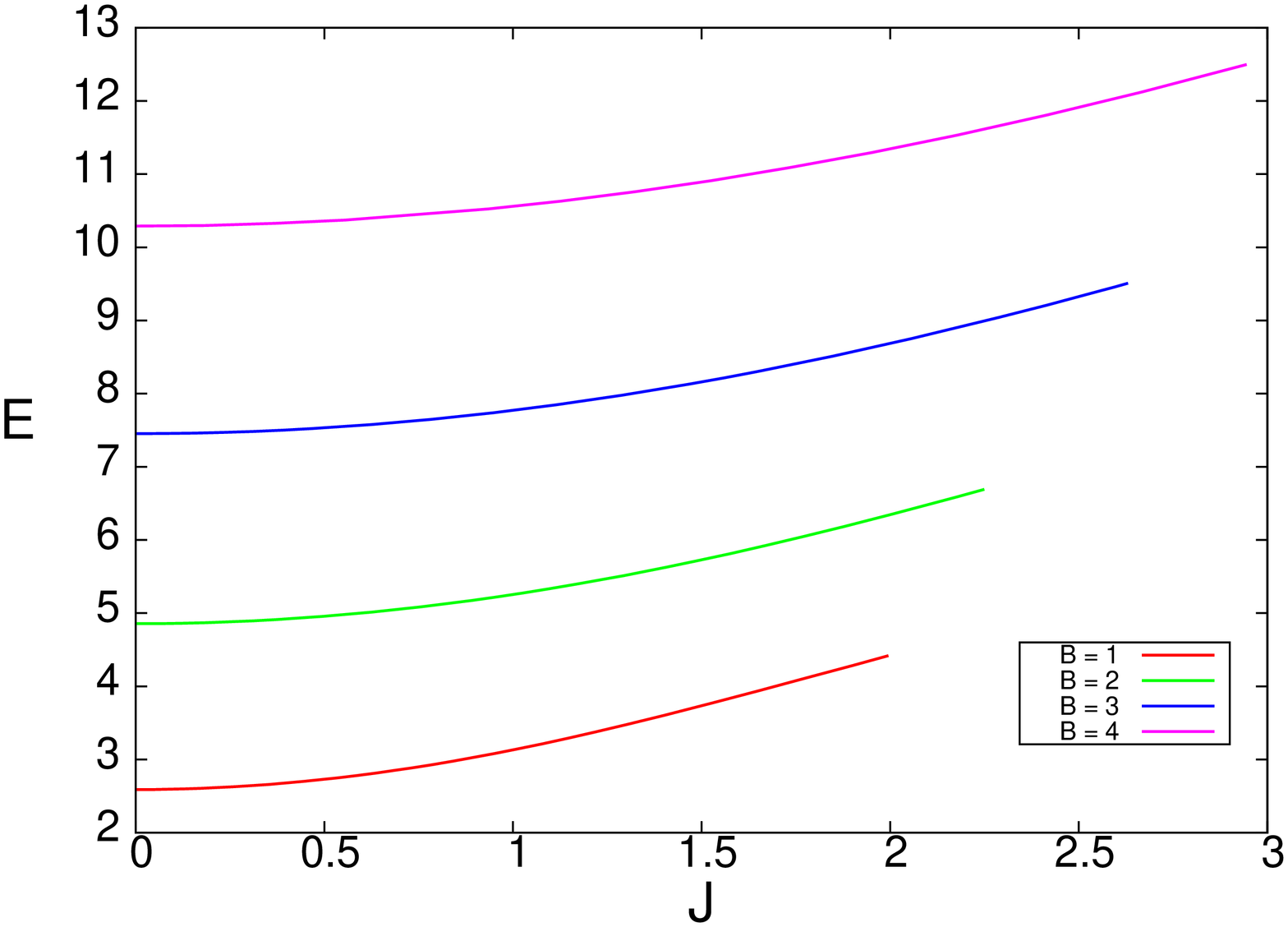}}
\end{center}
\caption{(Color online) Pseudoenergy, energy and isospin of the rotationally invariant baby Skyrmion solutions in the model with potential \re{pot-old}
are plotted as functions of angular frequency $\omega$. The energy is presented as a function of isospin $J$ at $\mu^2=4$.}
\end{figure}
%%%%%%%%%%%%%%%%%%%%%%%%%%%%%%%%%%%%%%%%%%%%%%%%%%%%%%%%%%%%%%%%%%%

The violation of the rotation invariance in the baby Skyrme model attracted a lot of attention recently, it was demonstrated
that the effect strongly depends on the particular choice of the potential of the model \cite{Ward,Hen,JSS}. Thus, considering
the isorotating baby Skyrmions we will consider complete system of coupled partial differential equations on the triplet
of functions $\bfph(\rho,\theta)$ which follows from the Lagrangian \re{Lag} in two cases of the rotationally invariant
potentials, the standard potential \re{pot-old} and the double vacuum model \re{double}. Note that our numerical results indicate
that another possible choice of the rotationally invariant
holomorphic potential $U[\bfph] = \mu^2 [1-\phi_3]^4$ \cite{Leese} does not
admit isorotations, the configuration becomes unstable for any non-zero value of the parameter $\omega$.

\section{Numerical results}

In this section the results of numerical simulations of the isospinning baby Skyrme model will be presented. The numerical calculations are
mainly performed on a equidistant grid in polar coordinates $\rho$ and $\theta$, employing the compact radial coordinate $x= \rho/(1+\rho) \in
[0:1]$ and $\theta \in [0,2\pi]$. To find stationary points of the functional \re{F}, which depends parametrically on $\omega$,
we implement a simple forward differencing scheme on a square lattice with lattice spacing $\Delta x = 0.01$.
Typical grids used have sizes $120 \times 120$.
The relative errors of the solutions are of order of $10^{-4}$ or smaller. To check our results for correctness
we checked that the corresponding virial relation \re{virial} holds,
as a further verification we evaluated the value of the topological charge by direct integration of
\re{charge}.

Each of our simulations began at $\omega =0$ at fixed value of $\mu$,
then we proceed by making small increments in $\omega$. Initial configurations we created using ansatz
\be \label{rot-inv-ans}
\phi_1 = \sin f(\rho) \cos (B \theta);\quad
\phi_2 = \sin f(\rho) \sin (B \theta);\quad
\phi_1 = \cos f(\rho)
\ee
where the input profile function is defined as $f(\rho)=4 \arctan e^{-\rho}$. Evidently this corresponds to the configuration of
degree $B$ with standard boundary conditions on the profile function $f(\rho)$.
Generally, in our calculations
we do not impose any assumptions about the spatial symmetries of the components of the field $\bfph$ although in
the next section we briefly considered the axially-symmetric isospinning multiSkyrmions.

\subsection{Old Baby Skyrme model}
First we consider evolution of the baby Skyrmions in the model with potential \re{pot-old}.
Note that the rotationally invariant ansatz \re{rot-inv-ans}
corresponds to the ring baby Skyrmions which in the standard conventions, for $B>2$ and $\mu^2 = 0.1$
are not absolute minima but rather the saddle points of the energy functional \cite{Bsk,Foster}, they are unstable with respect to
perturbations which break this symmetry\footnote{We remind that the kinetic term in standard conventions
differs from our term by a factor of one-half.}.
However the situation may change as the mass parameter increases, for larger values of $\mu$
the solutions may preserve the axial symmetry. Indeed the larger parameter $\mu$ is the
smaller the soliton size is, in the limit of very large mass the potential term can be considered as a sort of constraint on
the field component $\phi_3$ with $\mu^2$ acting like a Lagrange multiple. Note that in that limit the baby Skyrmions  are, in fact,
compactons, the fields reach the vacuum values at finite distance from the center of the soliton and they do not have
asymptotic tails \cite{Adam:2009px}.
In our calculations we mainly considered relatively large values of $\mu >1$.

In Fig. \ref{f-2} we present typical graphs
of the soliton energies, both as functions of $\omega$ and as functions of isospin $J$ for a range of values of $\mu$.
When the mass parameter $\mu^2 < 2$ we observe critical behavior of the first type, the effective potential vanishes
and both the energy and the angular momentum diverge. When $\mu^2$ increases further, second type of
critical behavior is observed, our algorithm ceases to find any critical points when $\omega$ is taking the values
$\omega > \sqrt 2$
thought the energy and the angular momentum remain finite. Note that the plots of the energy of the baby Skyrmions
as function of isospin look similar with the dependencies $E(J)$ in the Faddeev--Skyrme model \cite{JHSS},
up to some value of $J$
the energy remains almost constant, i.e. the configuration spins as a rigid rotator, then the curve $E(J)$
becomes linear up to
critical value at which the solution breaks up.

%%%%%%%%%%%%%%%%%%%%%%%%%%%%%%%%%%%%%%%%%%%%%%%%%%%%%%%%%%%%%%%%%%%%%%%%%
\begin{figure}[tbh]
\begin{center}
\setlength{\unitlength}{1cm}
\lbfig{f-1}
\begin{picture}(0,12.0)
\put(-9.0,11.8)
{\mbox{
\psfig{figure=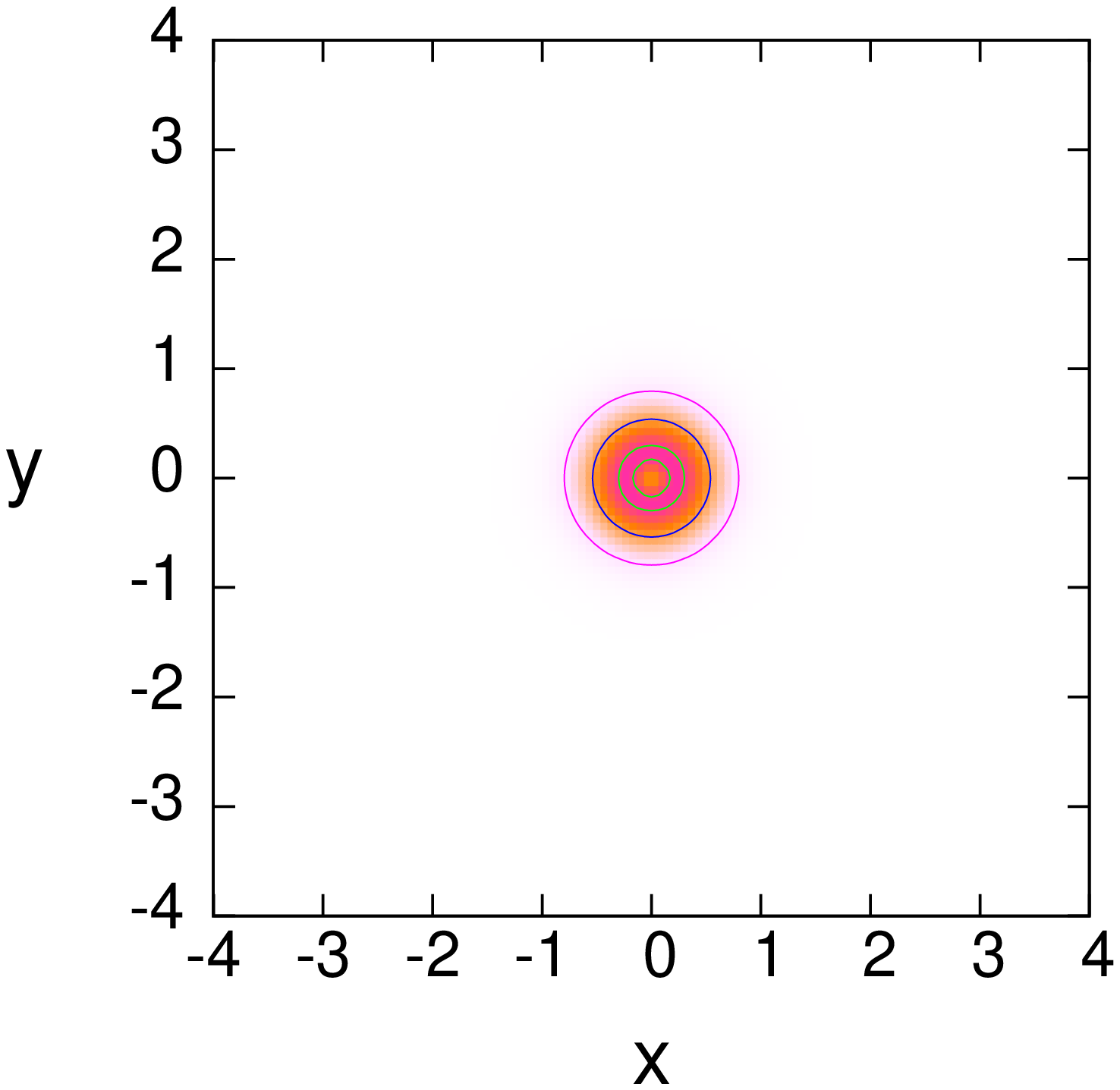,height=5.4cm, angle =-90}}}
\end{picture}
\begin{picture}(0,0.0)
\put(-5.0,11.8)
{\mbox{
\psfig{figure=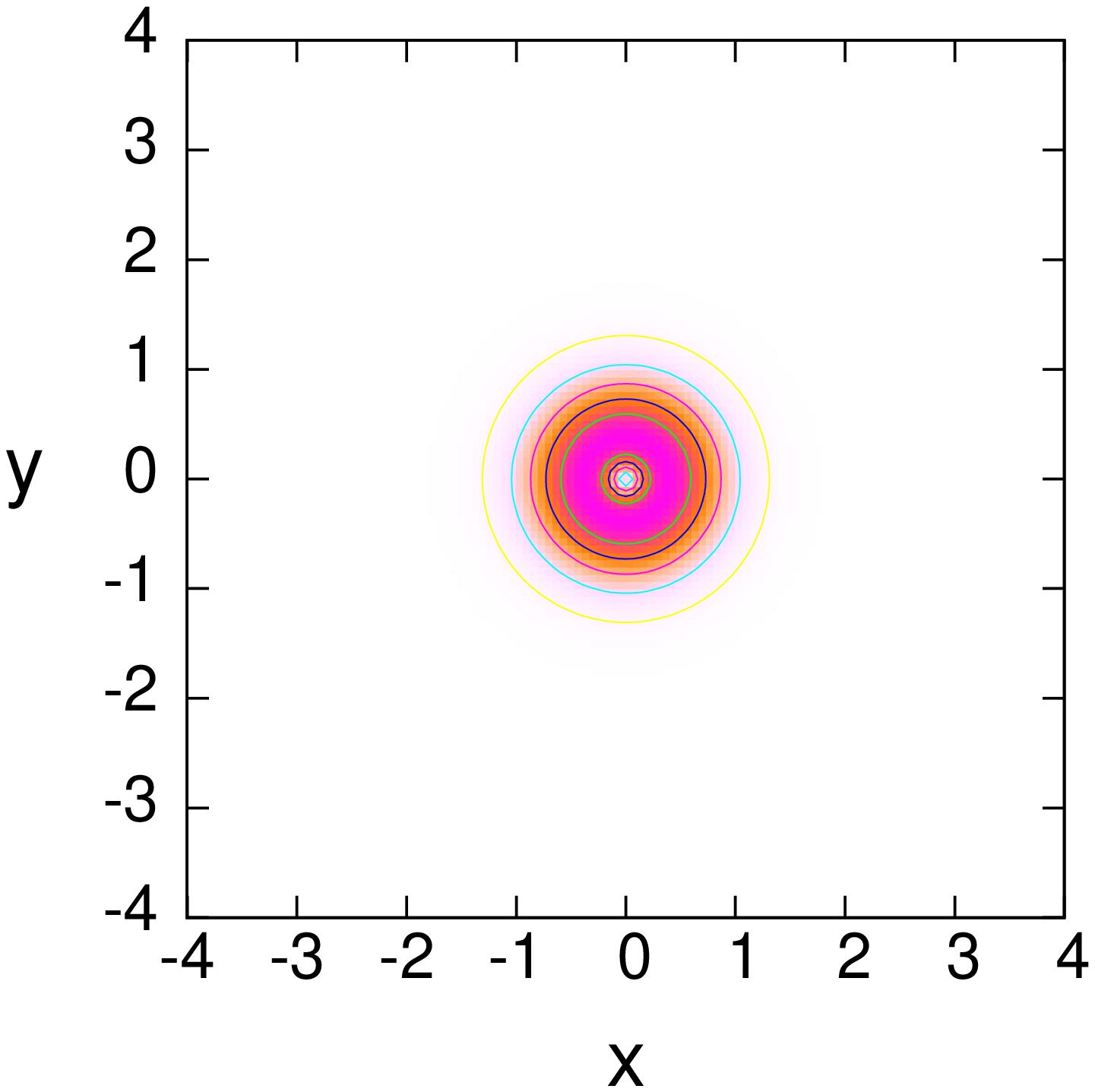,height=5.4cm, angle =-90}}}
\end{picture}
\begin{picture}(0,0.0)
\put(-1.2,11.8)
{\mbox{
\psfig{figure=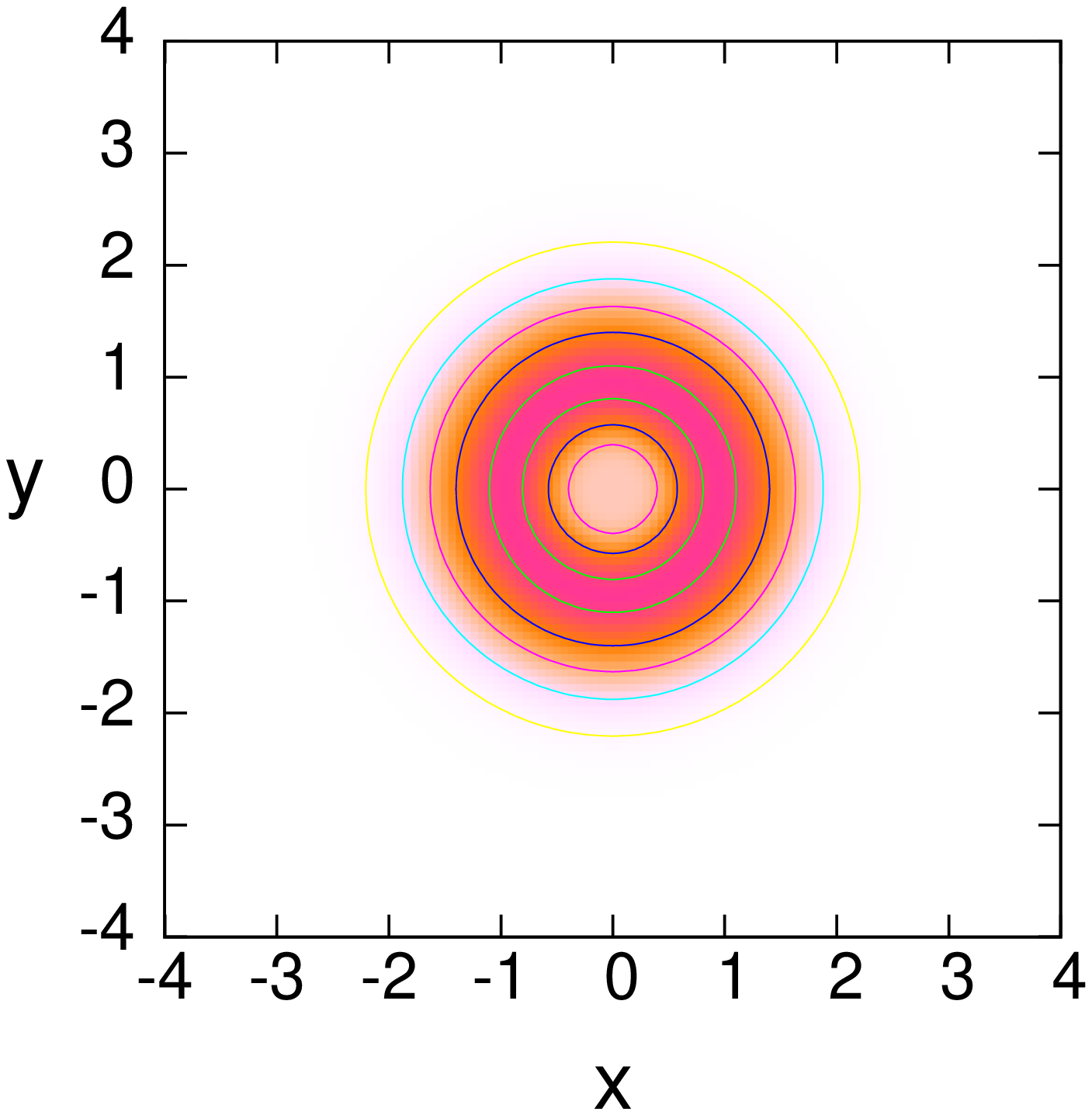,height=5.4cm, angle =-90}}}
\end{picture}
\begin{picture}(0,0.0)
\put(2.8,11.8)
{\mbox{
\psfig{figure=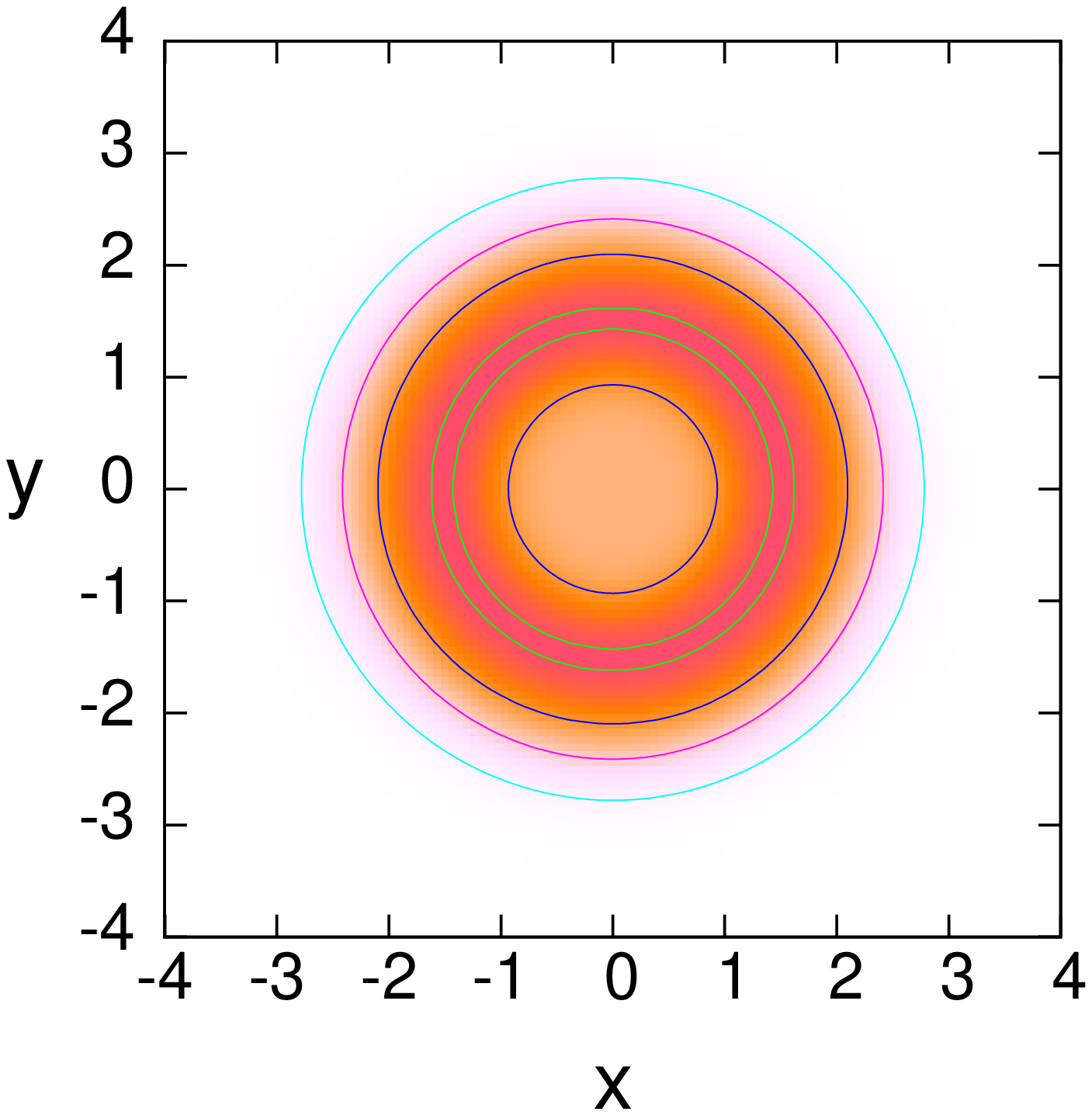,height=5.2cm, angle =-90}}}
\end{picture}
\begin{picture}(0,0.0)
\put(-9.5,7.8)
{\mbox{
\psfig{figure=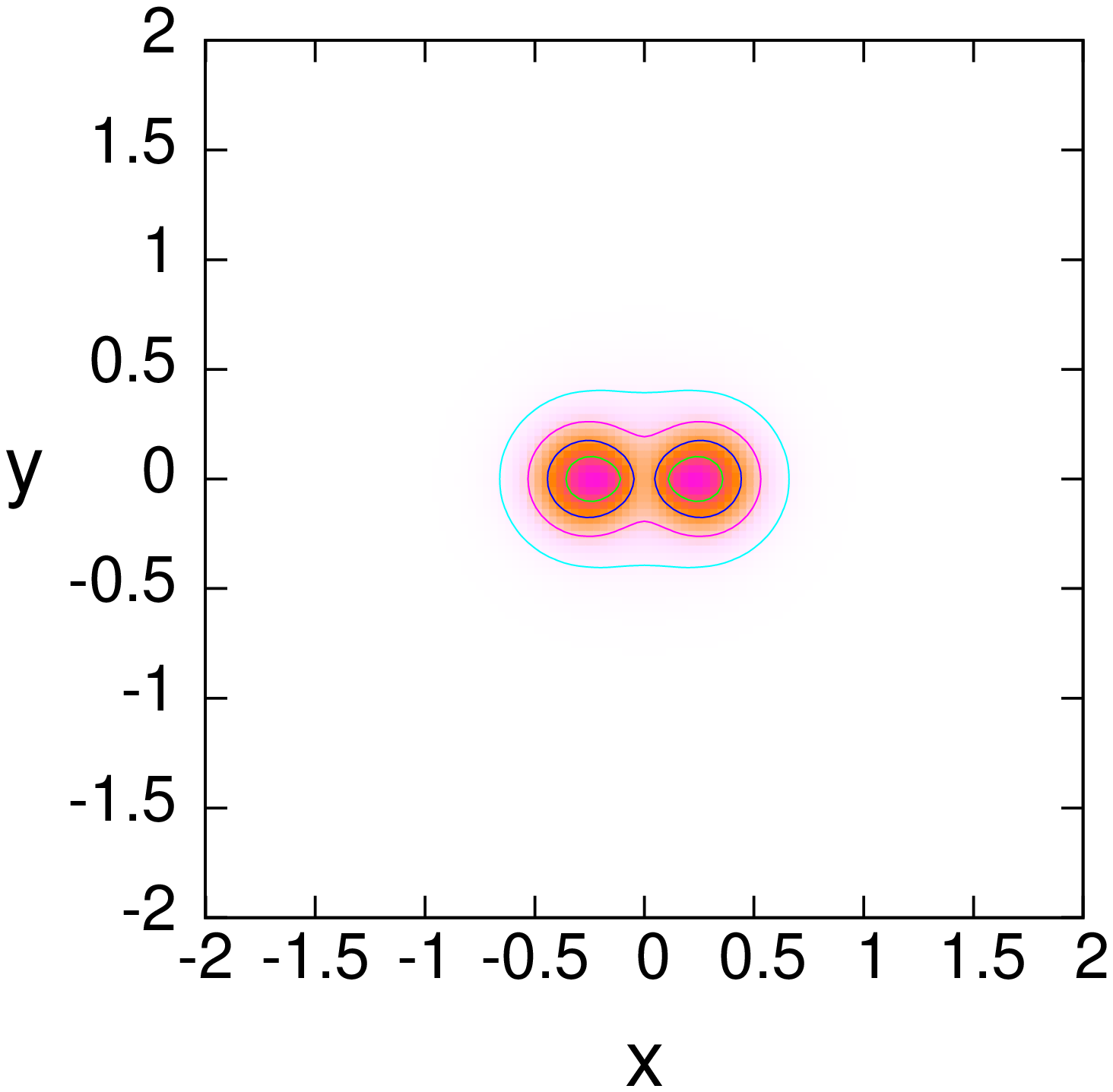,height=5.2cm, angle =-90}}}
\end{picture}
\begin{picture}(0,0.0)
\put(-5.5,7.8)
{\mbox{
\psfig{figure=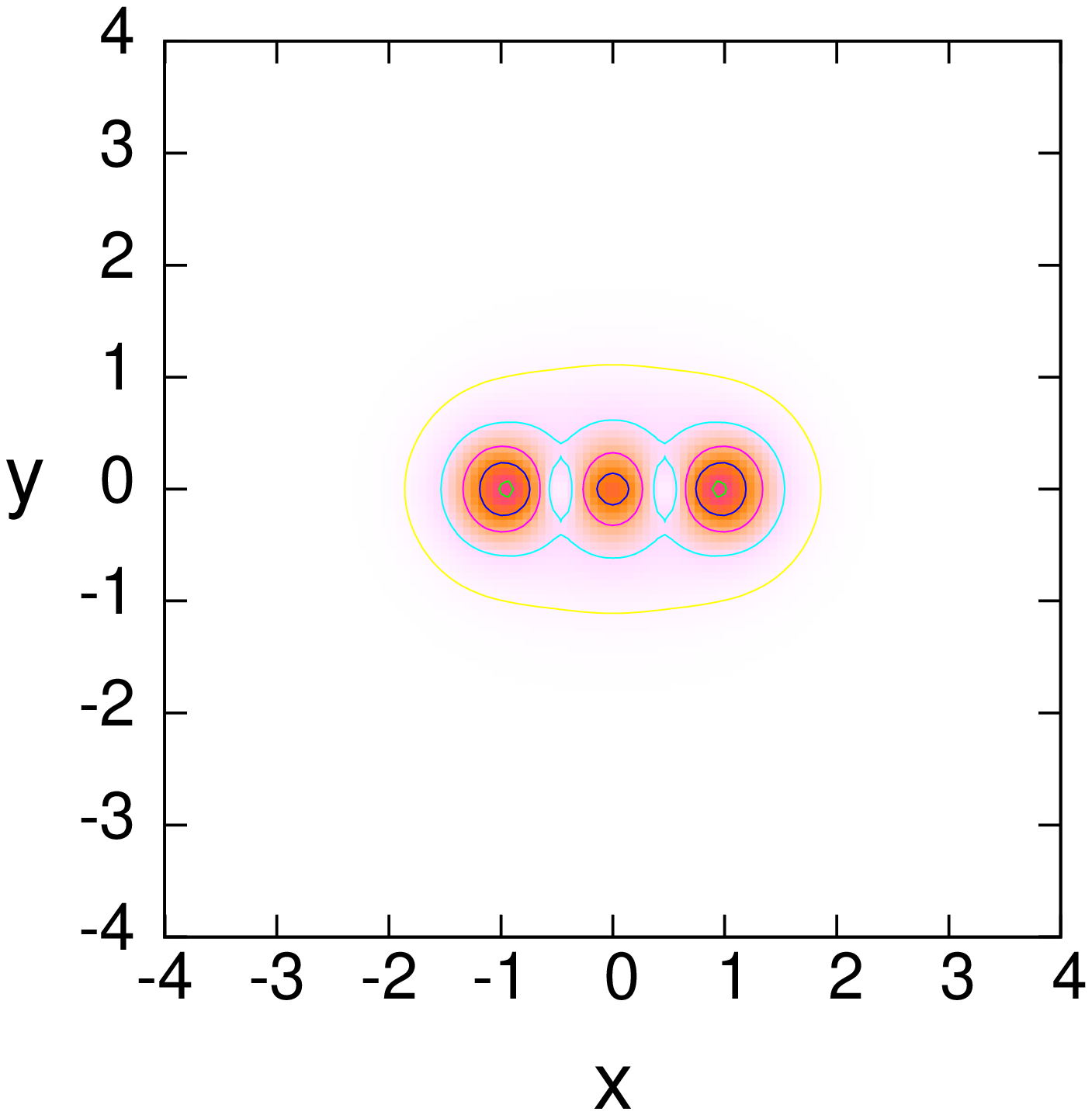,height=5.2cm, angle =-90}}}
\end{picture}
\begin{picture}(0,0.0)
\put(-1.6,7.8)
{\mbox{
\psfig{figure=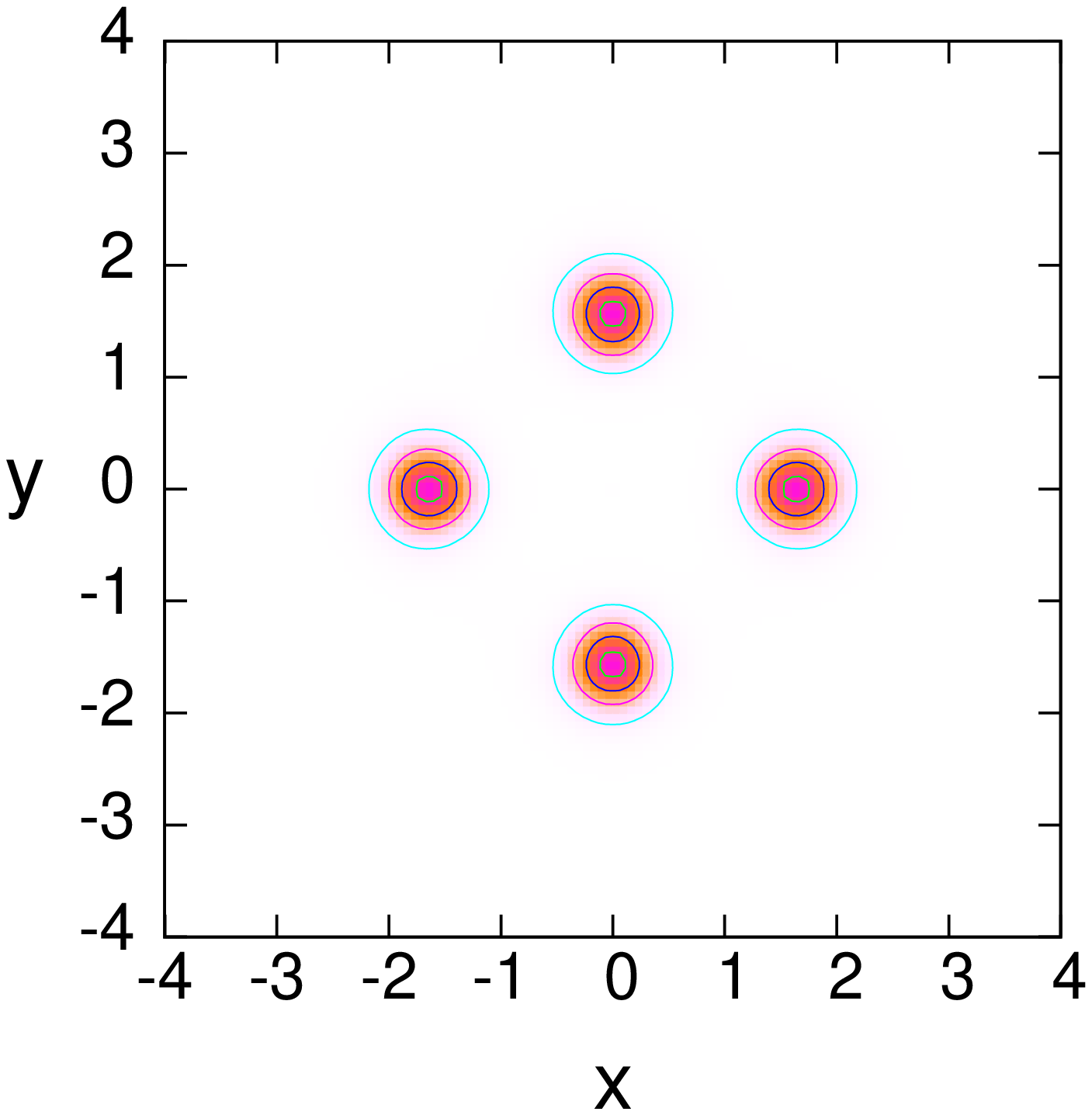,height=5.2cm, angle =-90}}}
\end{picture}
\begin{picture}(0,0.0)
\put(2.4,7.8)
{\mbox{
\psfig{figure=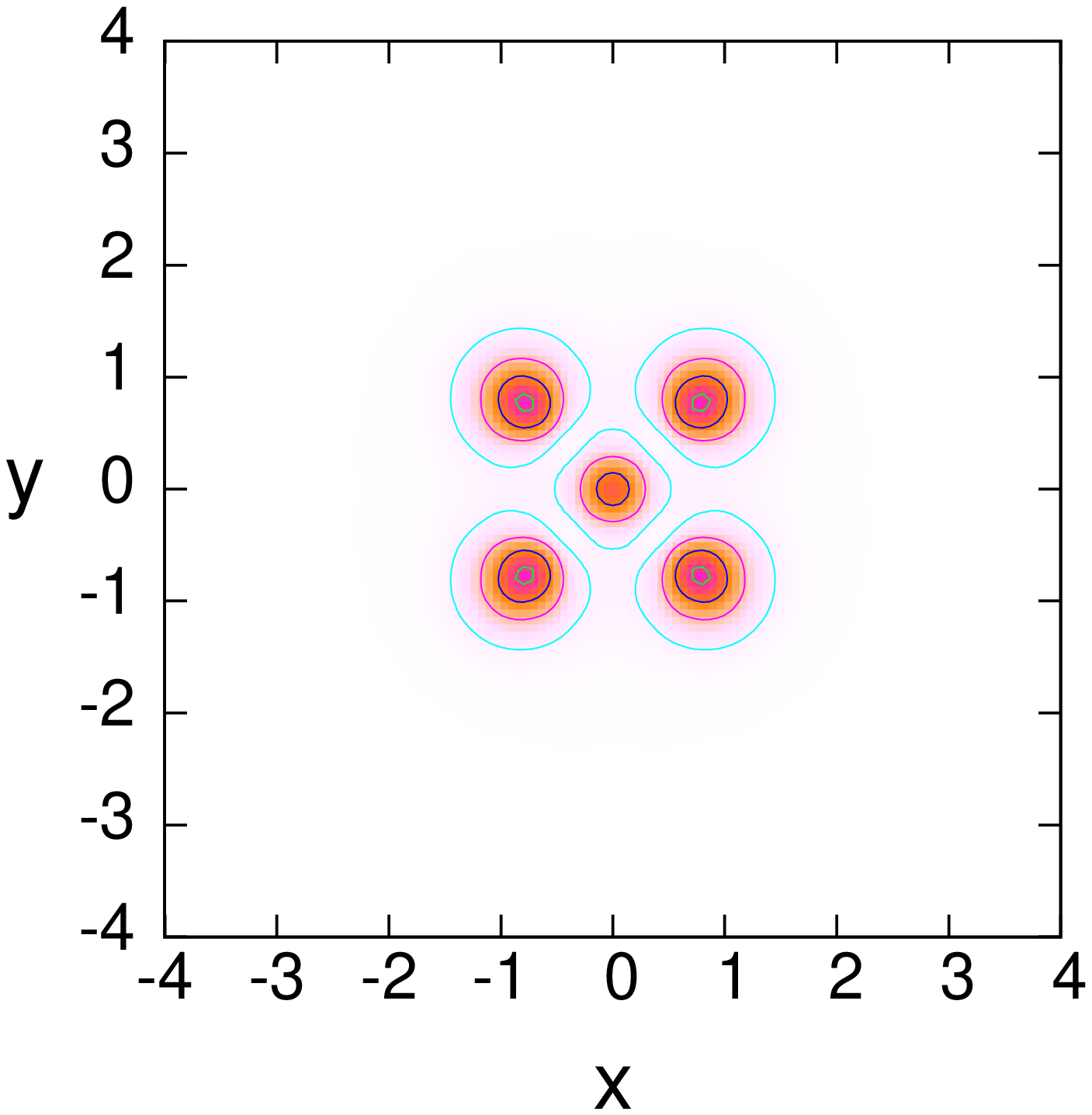,height=5.2cm, angle =-90}}}
\end{picture}
\begin{picture}(0,0.0)
\put(-9.5,4.1)
{\mbox{
\psfig{figure=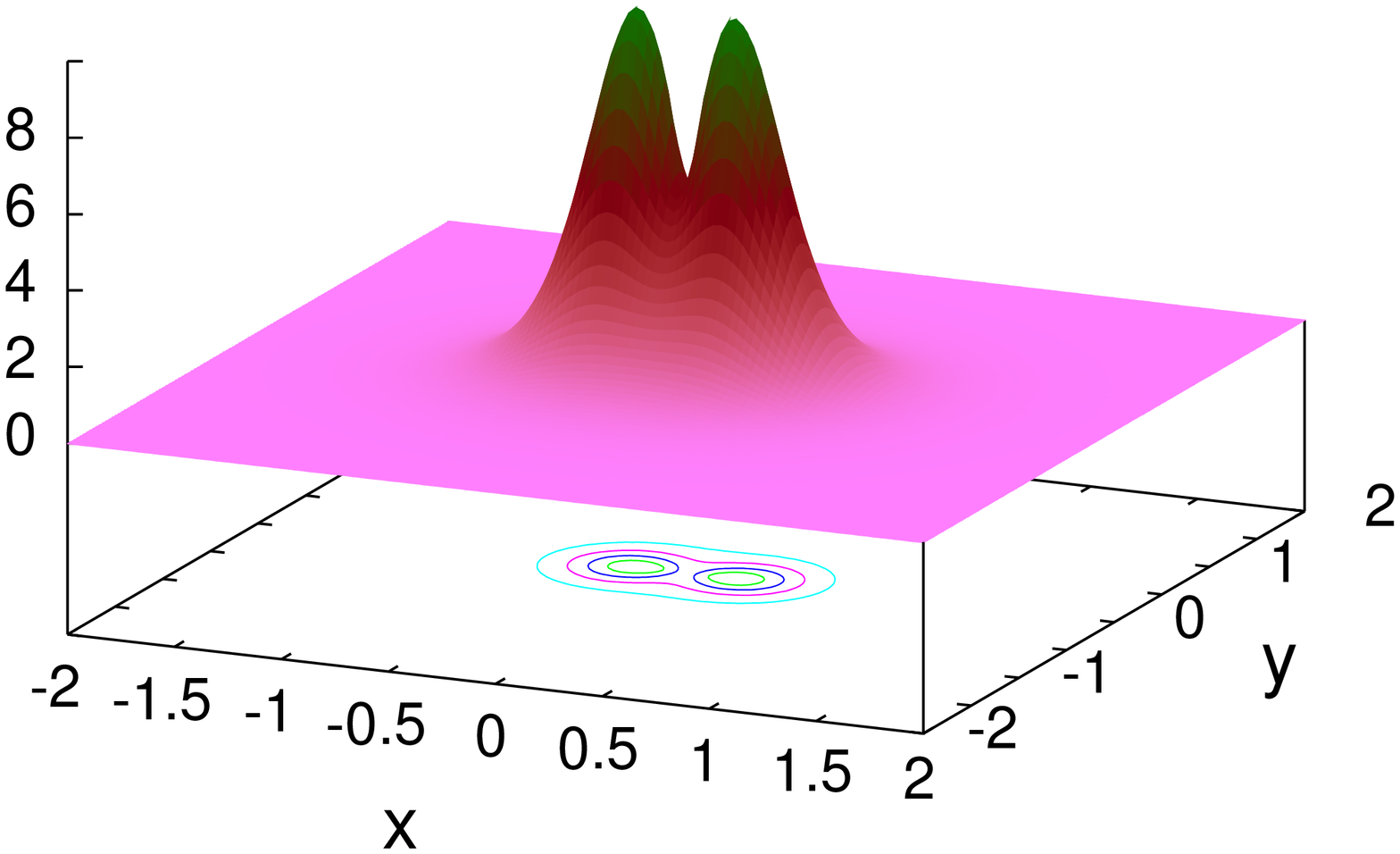,height=5.2cm, angle =-90}}}
\end{picture}
\begin{picture}(0,0.0)
\put(-5.5,4.1)
{\mbox{
\psfig{figure=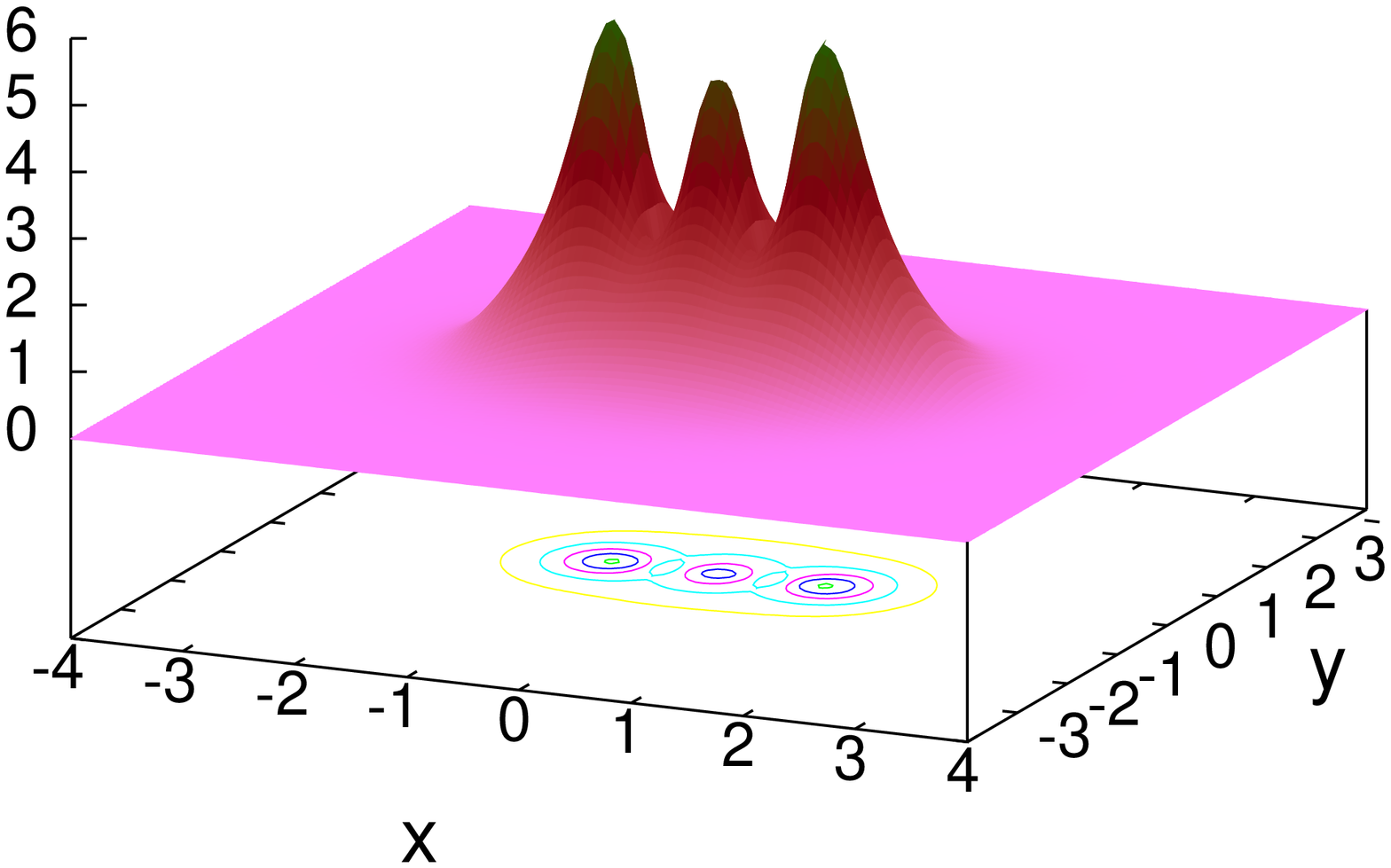,height=5.2cm, angle =-90}}}
\end{picture}
\begin{picture}(0,0.0)
\put(-1.6,4.1)
{\mbox{
\psfig{figure=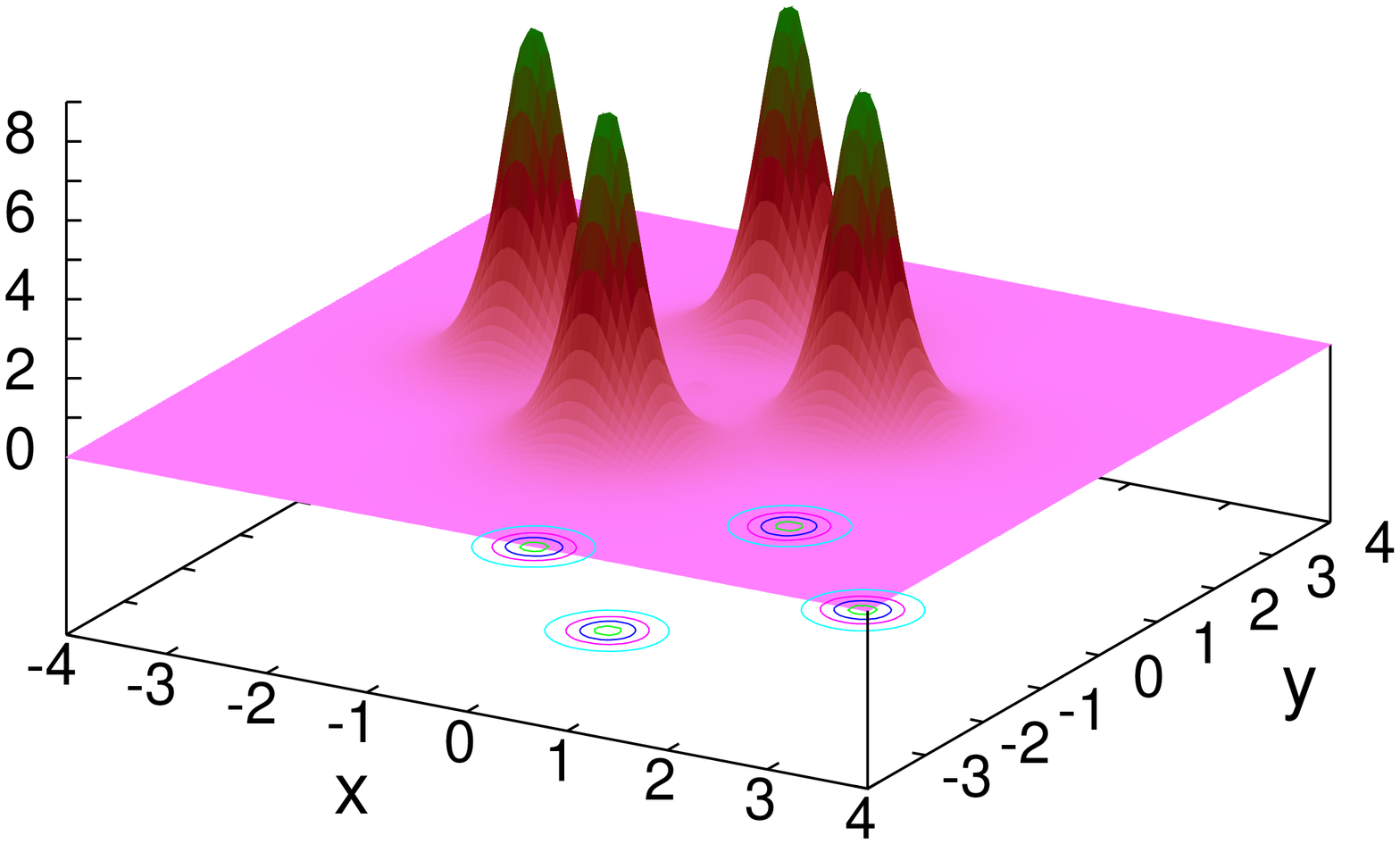,height=5.2cm, angle =-90}}}
\end{picture}
\begin{picture}(0,0.0)
\put(2.4,4.1)
{\mbox{
\psfig{figure=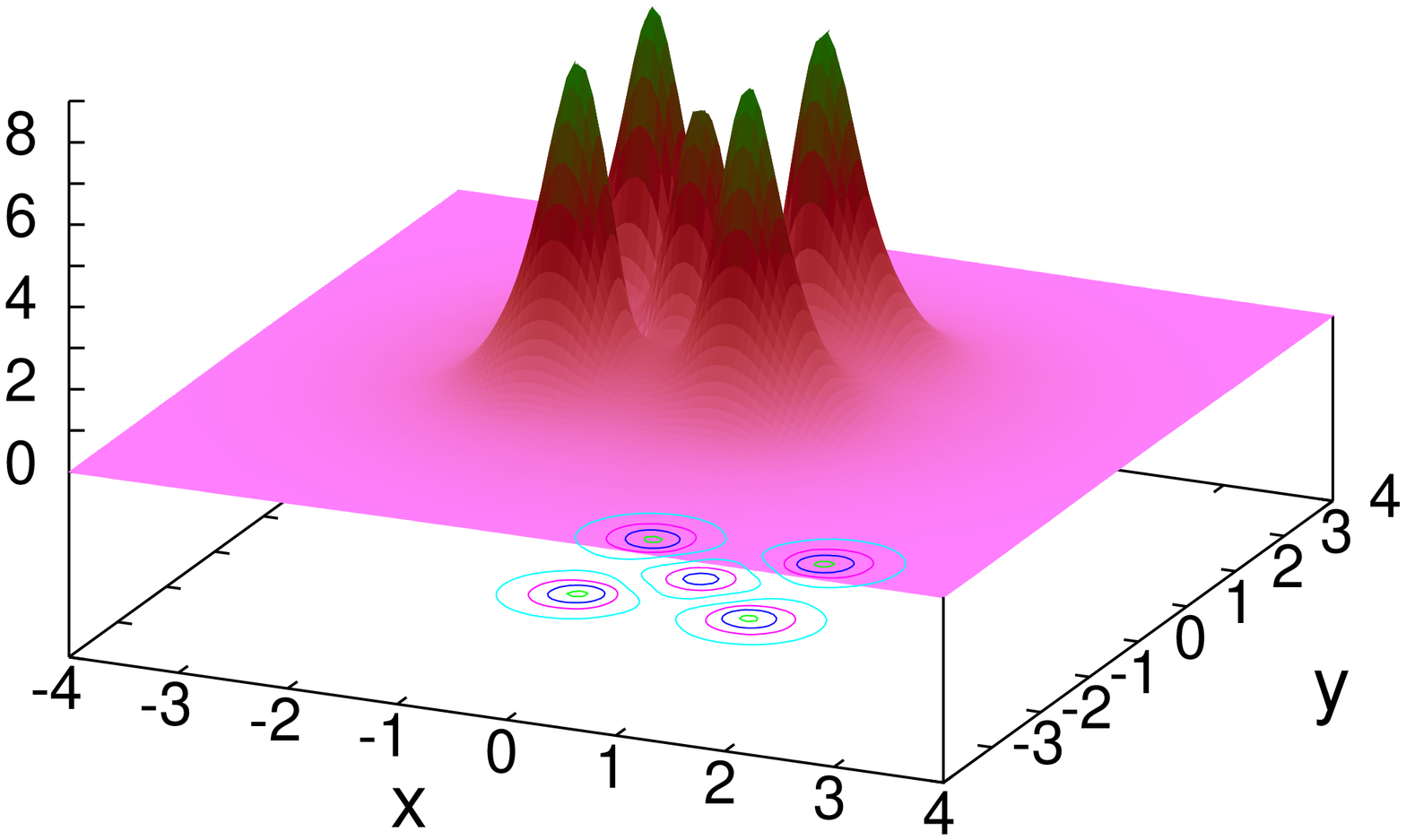,height=5.2cm, angle =-90}}}
\end{picture}
\end{center}
\caption{\small (Color online)
Critical behavior of the rotationally invariant soliton solutions of the model with potential \re{pot-old}.
The contour plots of the energy density of the rotationally invariant (upper row) baby Skyrmions with charges $B=2,3,4,5$
and $\mu^2=8$ at $\omega=0.8$ and their decay into $B$ charge one solitons (2nd and 3rd rows).}
\end{figure}
%%%%%%%%%%%%%%%%%%%%%%%%%%%%%%%%%%%%%%%%%%%%%%%%%%%%%%%%%%%%%%%%%%%%%%%%%

Interestingly, for the rotationally invariant configurations which we can construct using the 'hedgehog' ansatz \cite{Bsk} and considering
relatively large values of the mass parameter $\mu$,
we observe crossing in both $F_\omega (\omega)$ and $E(\omega)$ curves as displayed
on Fig.~\ref{f-3}. Indeed, our numerical simulations confirm that for
some (third) critical value of frequency $\omega_3$ the pseudo-energy of the axially symmetric $B\ge 2$
multi-Skyrmion becomes higher than the pseudo-energy of the system of $B$ charge one baby Skyrmions, so the
configurations are unstable with respect to decay into constituents as shown in Fig.~\ref{f-1}.
Typically, increasing the value of the mass parameter
$\mu$ will increase the stability of the rotationally invariant multisolitons, the critical values of the frequencies which
correspond to the crossing between the $F_\omega (\omega)$ curves then increase.
%It is known, however, that
%the axial symmetry of the non-rotating configurations of higher degree than two becomes broken \cite{PZS}, thus for these
%configurations we do not observe crossings in the $F_\omega (\omega)$ and $E(\omega)$ curves.

\begin{figure}
\lbfig{f-6}
\begin{center}
{\includegraphics[scale=0.24,angle=-90]{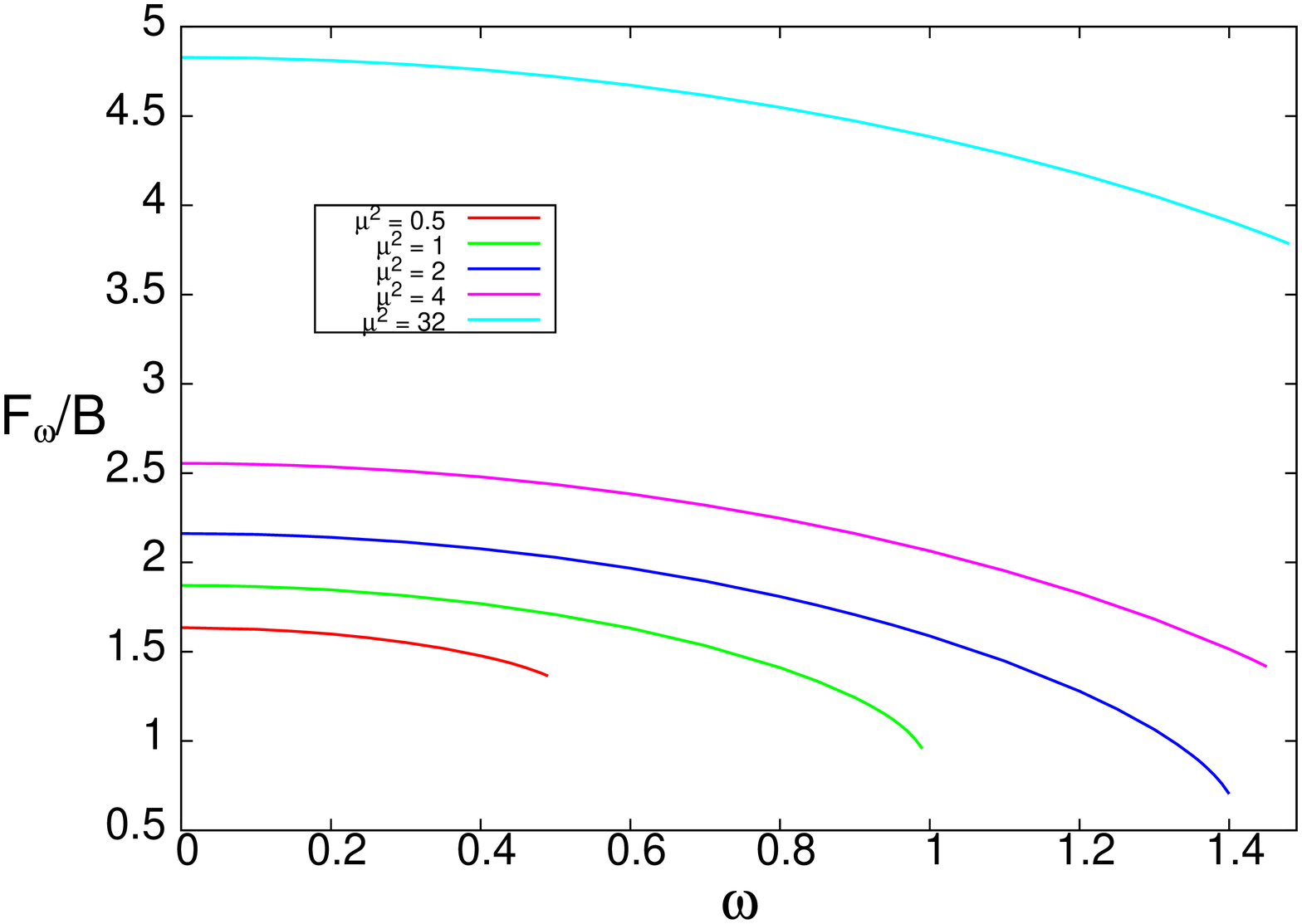}}
{\includegraphics[scale=0.24,angle=-90]{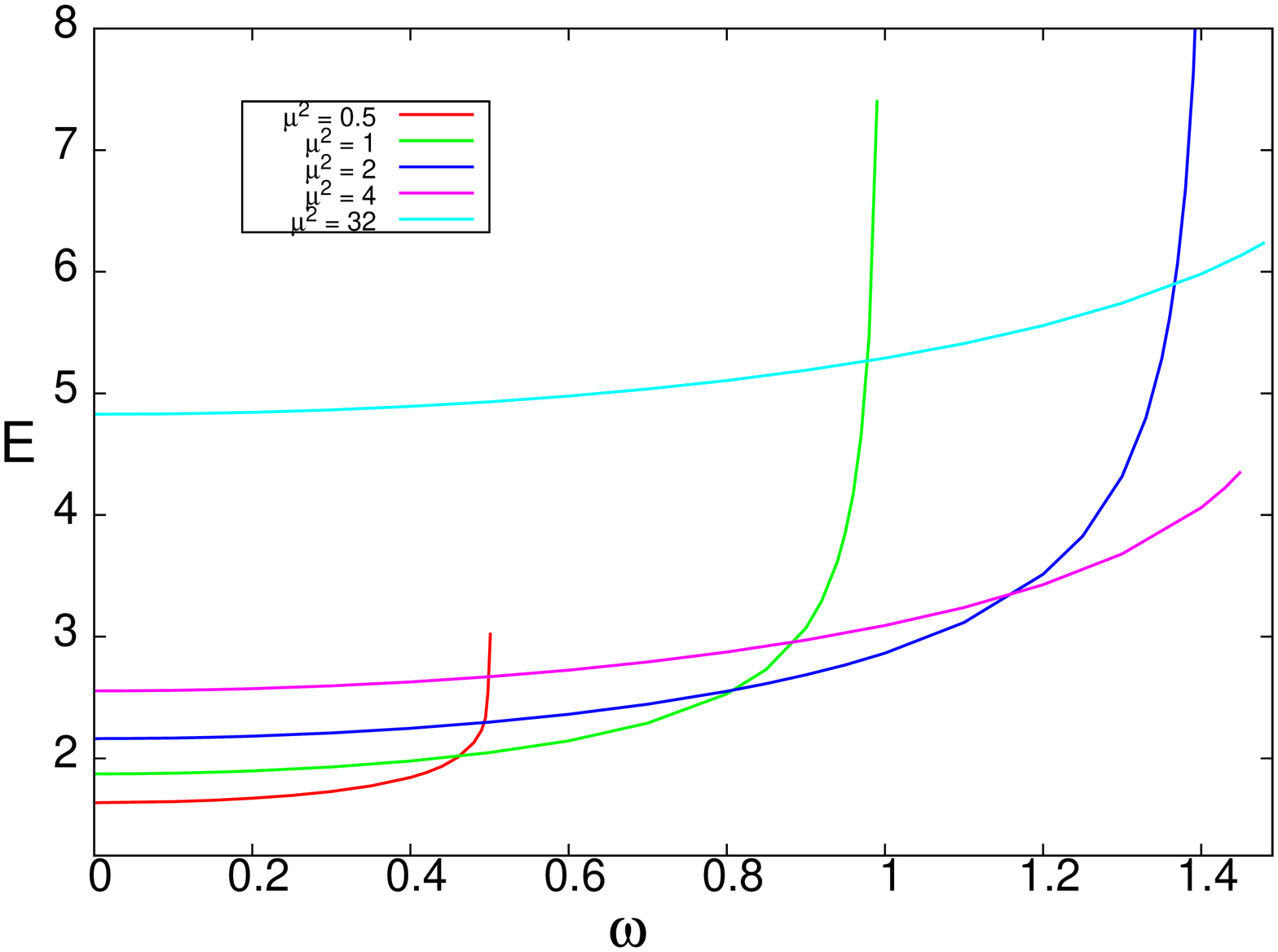}}
{\includegraphics[scale=0.24,angle=-90]{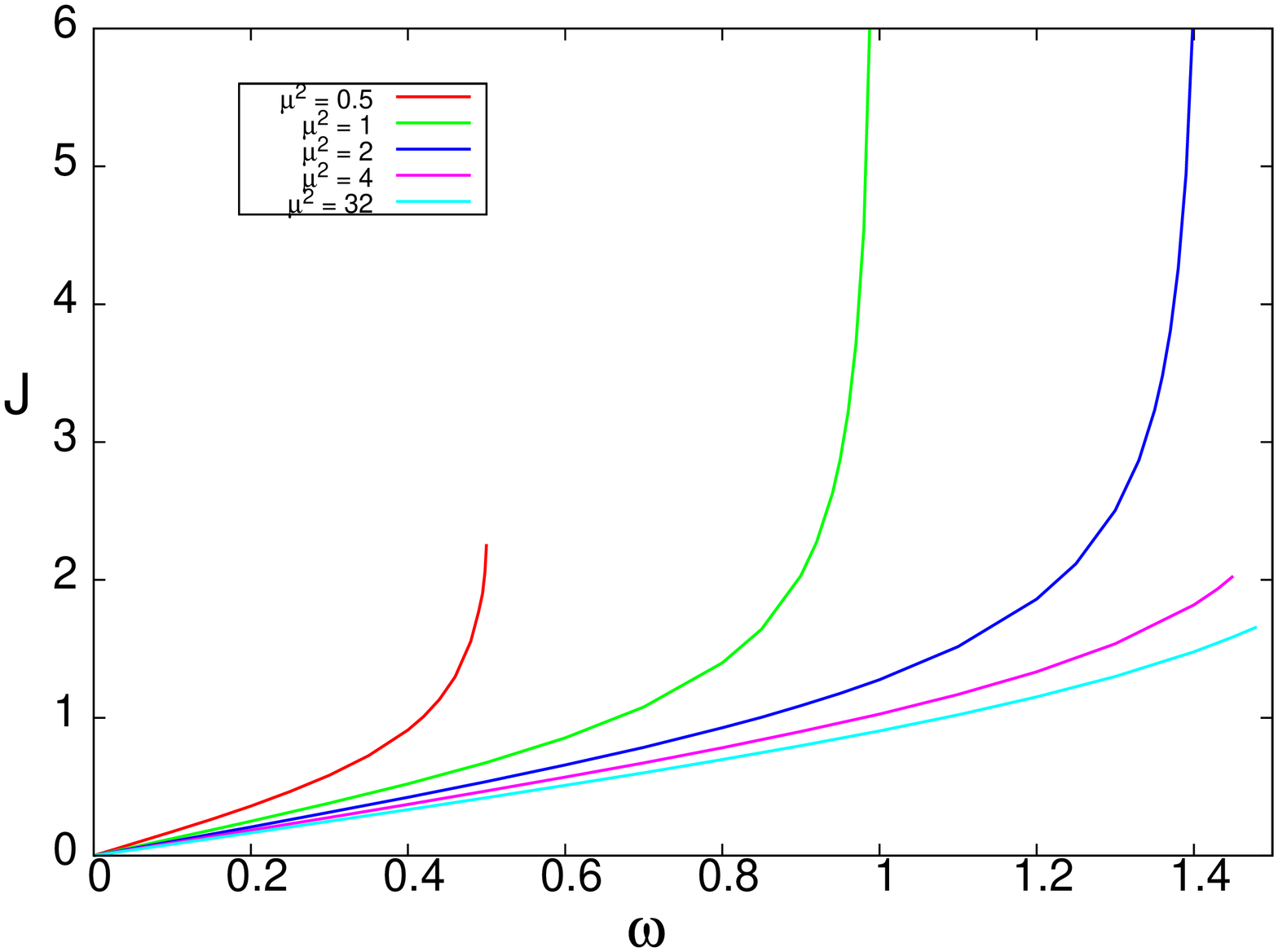}}
{\includegraphics[scale=0.24,angle=-90]{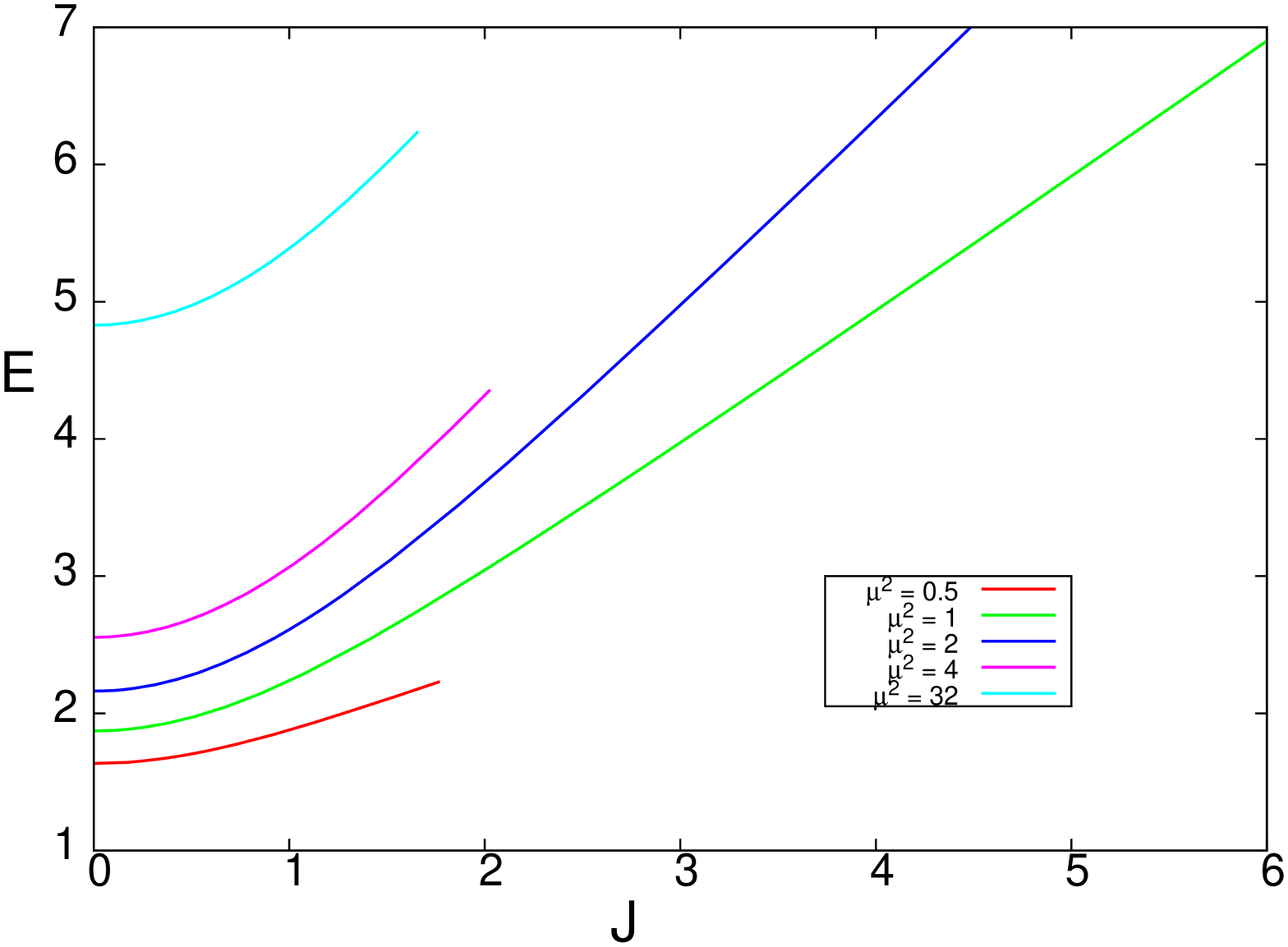}}
\end{center}
\caption{(Color online) Pseudoenergy, energy and isospin of the $B=1$ baby Skyrmion in the model with potential \re{double}
are plotted as functions of the angular frequency $\omega$. The energy $E$
is plotted as a function of isospin $J$ at $\mu^2=0.5,1,2,4,32$.}
\end{figure}

\subsection{Double vacuum Baby Skyrme model}
Let us now briefly consider the "new" baby Skyrme model with double vacuum potential \re{double}.
First, we observed some similarity with the
case of the model with old potential considered above. As the angular frequency $\omega$ increases the
radius of the configuration is getting larger whereas its qualitative shape does not change.
In Fig.~\ref{f-6}  we present typical plots
of the  $B=1$ Skyrmion energies, both as functions of $\omega$ and as functions of isospin $J$ for a range of values of $\mu$.
Again, we observe critical behavior of two different types,
if the mass parameter satisfies the condition $\mu^2 \le \omega_1^2 = 2$
the effective potential vanishes at $\mu = \omega_2$
and both the energy and the angular momentum diverge (cf Fig. \ref{f-2}).
In the second case  $\mu^2 > 2$ we observe another type of
the critical behavior, as $\omega$ approaches the critical value $\omega_1$ the rotational invariance of the configuration
becomes slightly broken as shown in Fig.~\ref{f-4} and then
the configuration cease to exist. However both the energy and the angular momentum of the Skyrmion remain finite up to
the critical value $\omega_1$.

Note that the charge $B$ solutions of the model with double
vacuum potential are rotationally invariant.
Both the pseudo-energy and the total energy of these
multi-Skyrmions  are always lower than the pseudo-energy and the energy of the system of $B$ individual charge one solitons,
so the system remains stable with respect to decay into constituents. Indeed, we
do not observe crossings in $F_\omega (\omega)$ curves displayed
on Fig.~\ref{f-5} which one has to compare with similar curves for the rotationally invariant soliton solutions
of the old baby Skyrme model we presented in Fig.~\ref{f-3}.

\begin{figure}
\lbfig{f-5}
\begin{center}
{\includegraphics[scale=0.24,angle=-90]{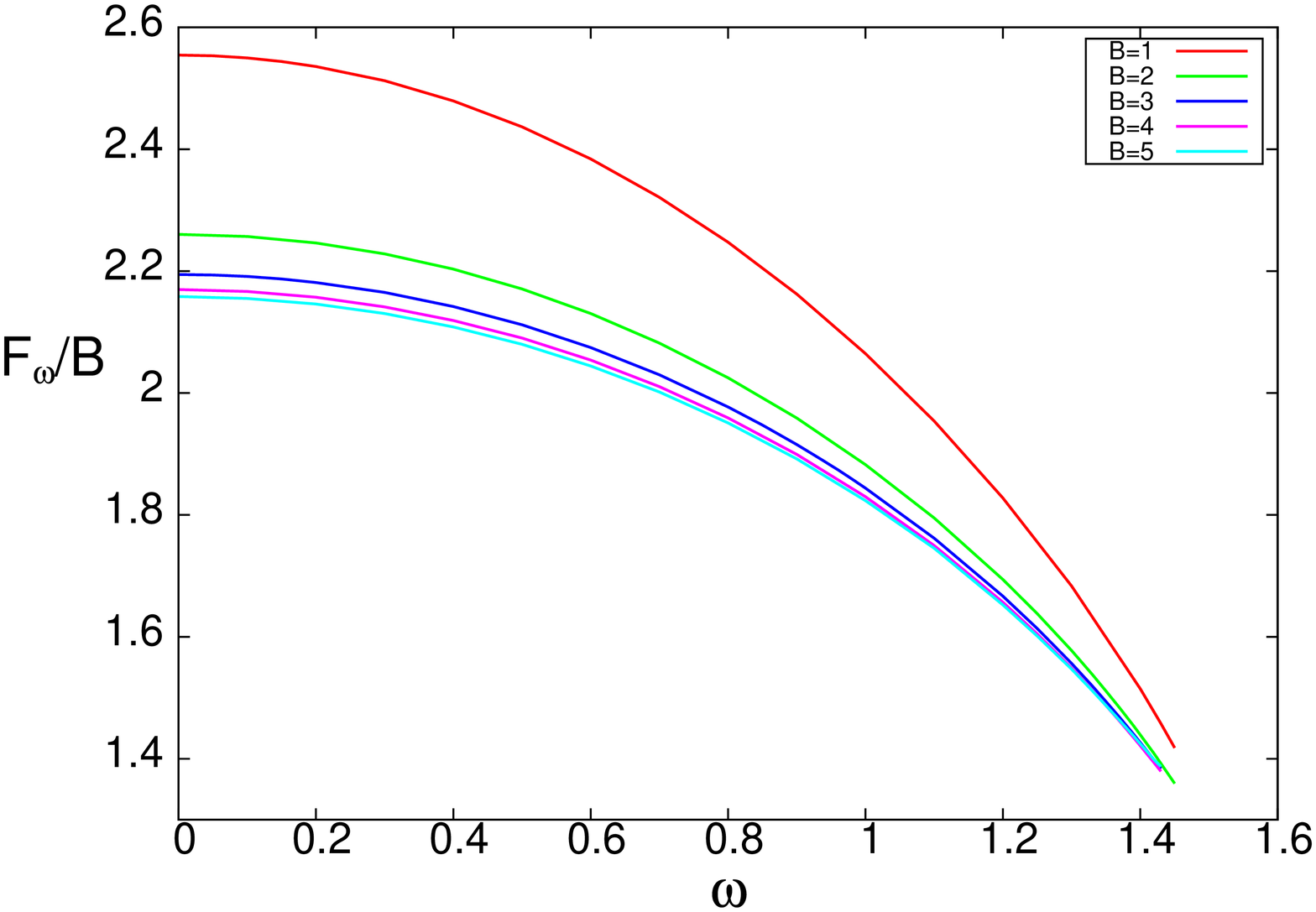}}
{\includegraphics[scale=0.24,angle=-90]{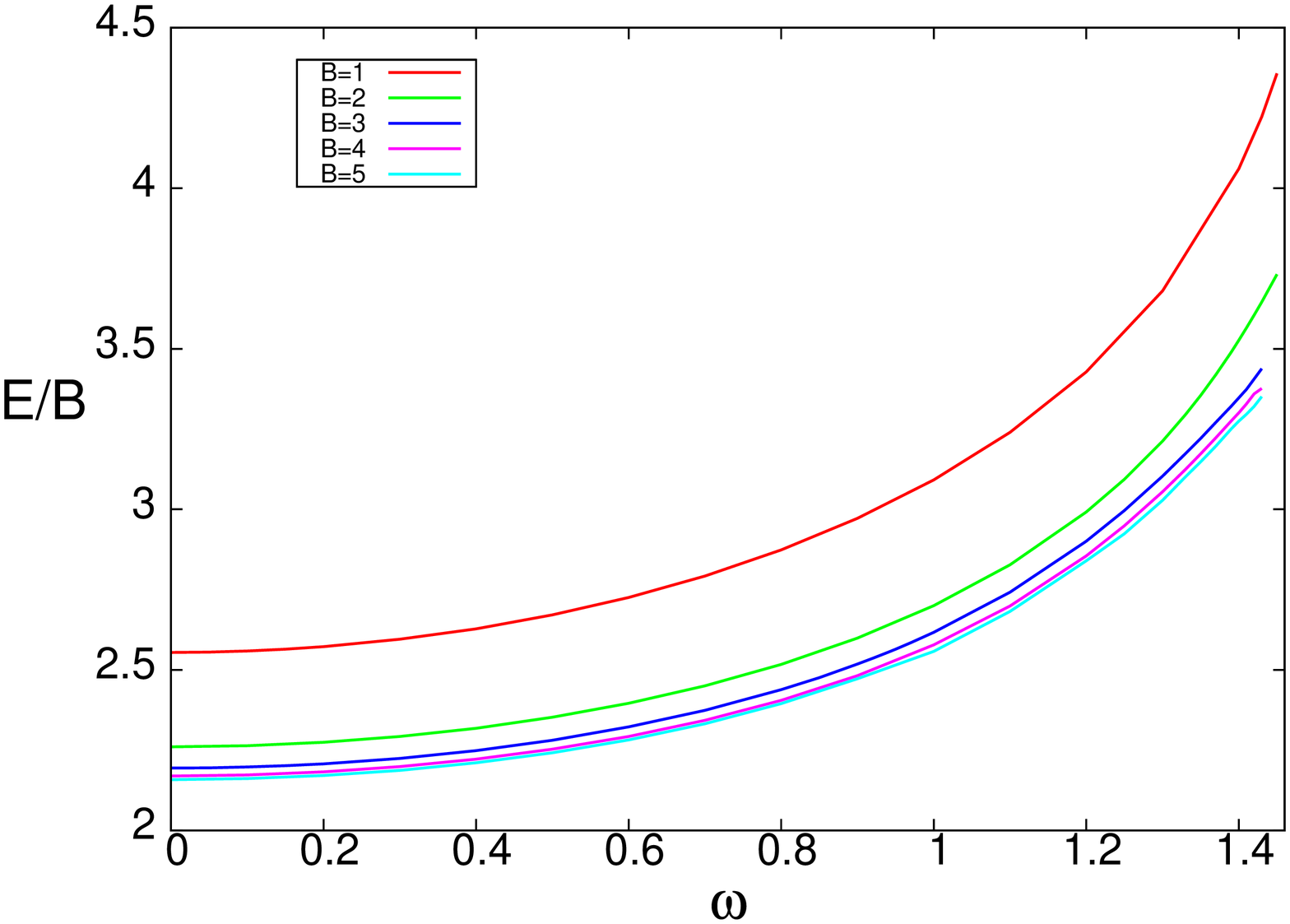}}
{\includegraphics[scale=0.24,angle=-90]{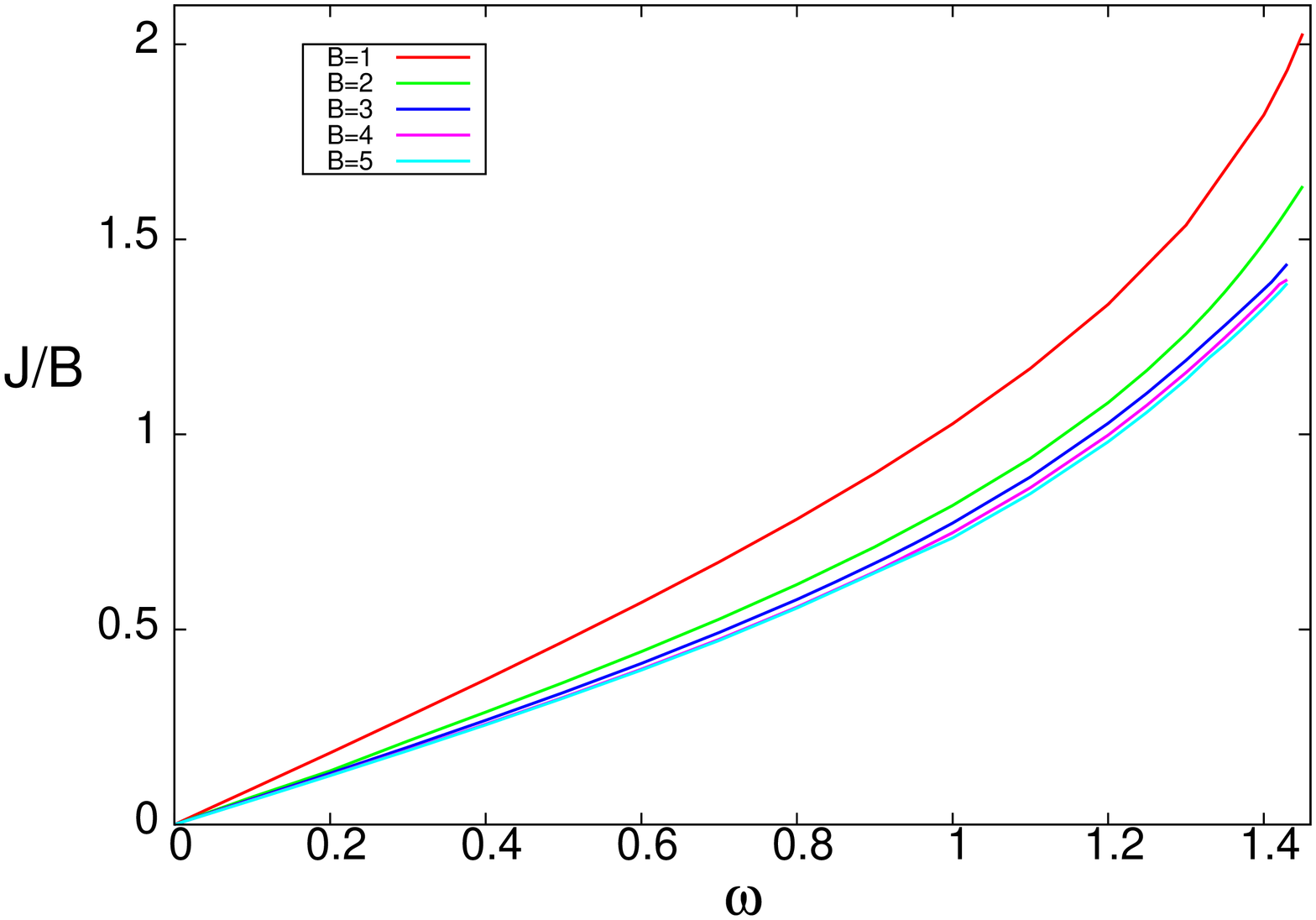}}
{\includegraphics[scale=0.24,angle=-90]{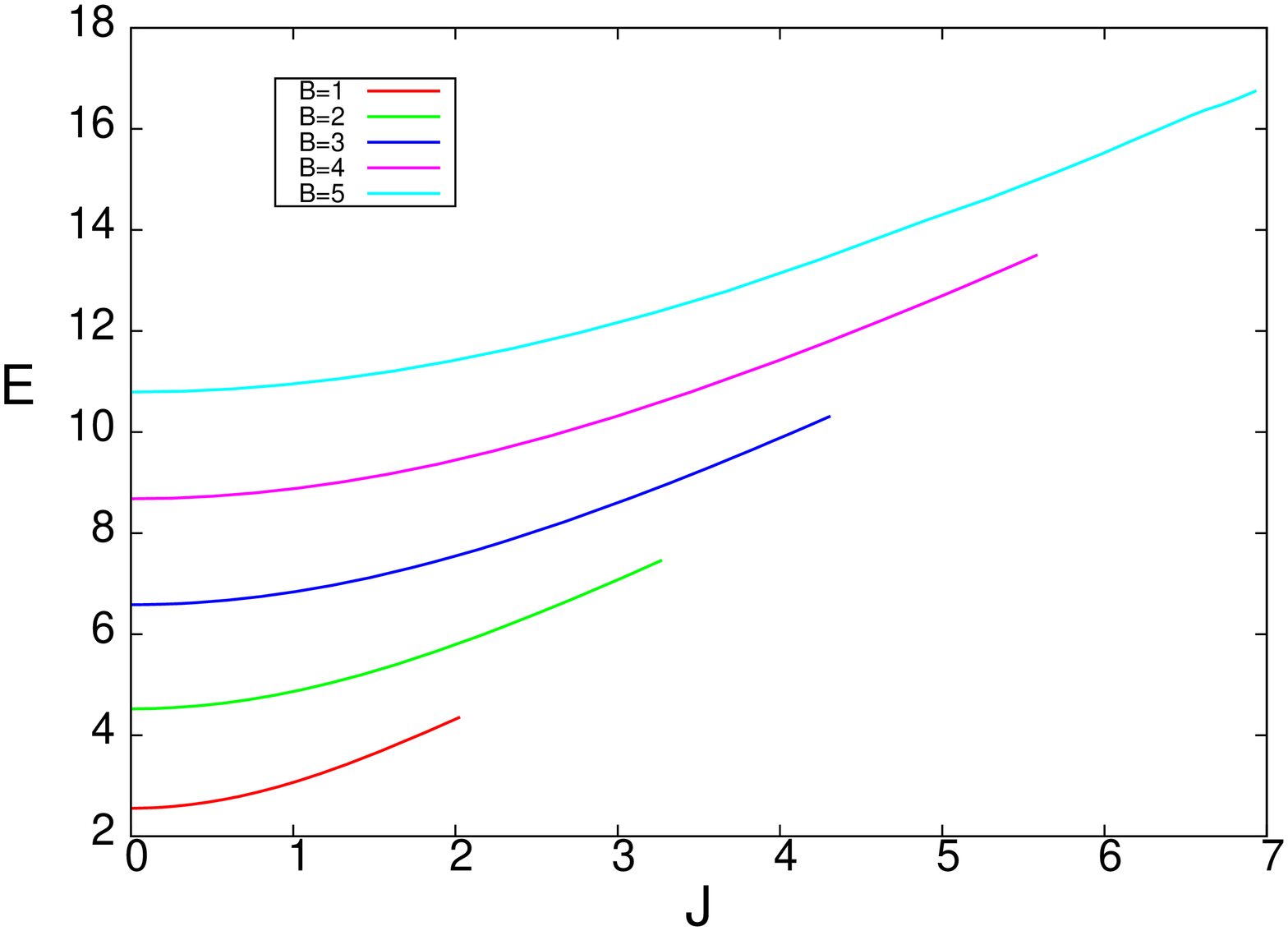}}
\end{center}
\caption{(Color online) Pseudoenergy, energy and isospin of the baby Skyrmion solutions in the model with potential \re{double}
are plotted as functions of angular frequency $\omega$, and the energy as a function of isospin $J$ at $\mu^2=4$.}
\end{figure}
%%%%%%%%%%%%%%%%%%%%%%%%%%%%%%%%%%%%%%%%%%%%%%%%%%%%%%%%%%%%%%%%%%%

\begin{figure}
\lbfig{f-4}
\begin{center}
{\includegraphics[scale=0.31,angle=-90]{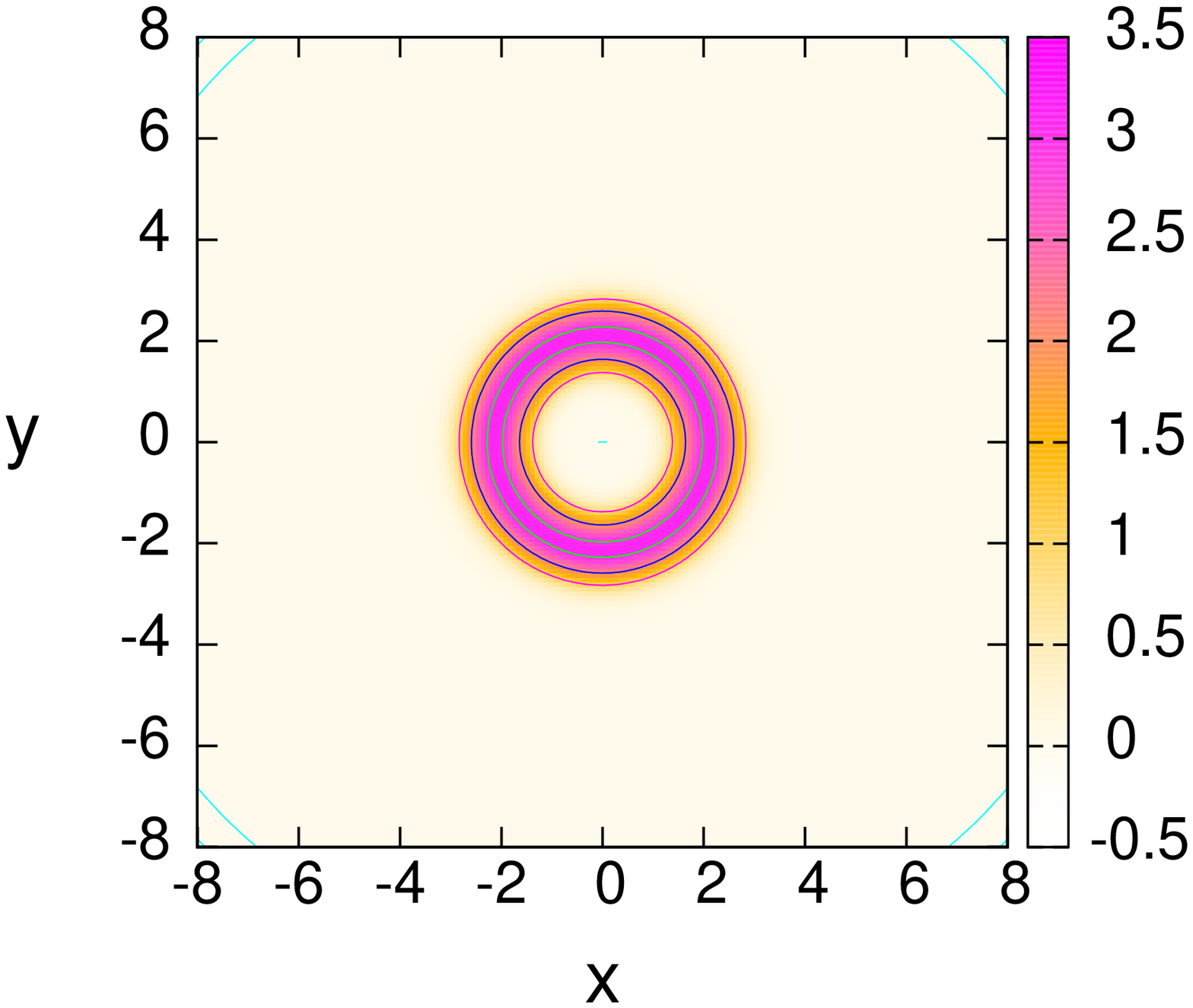}}
{\includegraphics[scale=0.31,angle=-90]{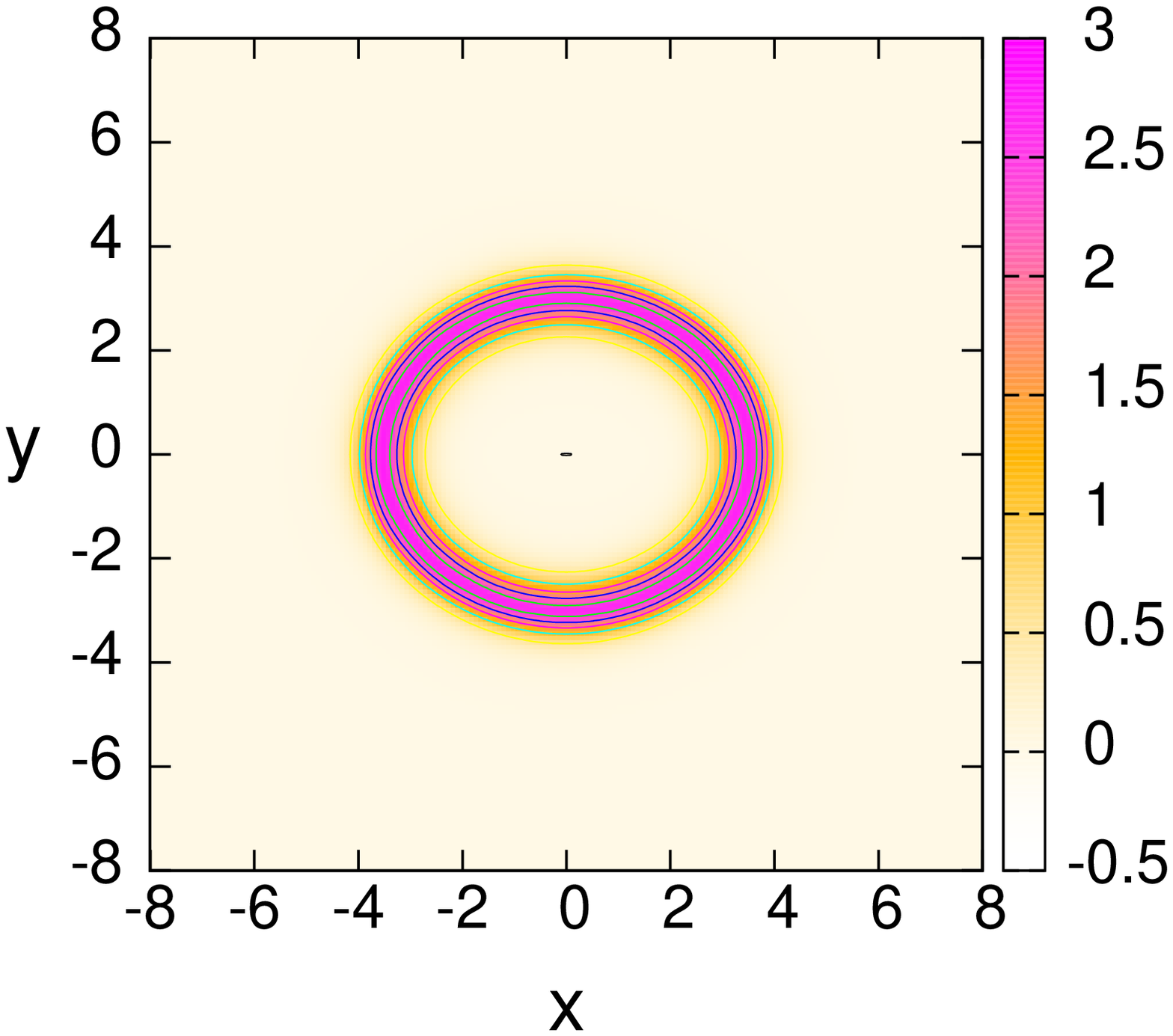}}
{\includegraphics[scale=0.31,angle=-90]{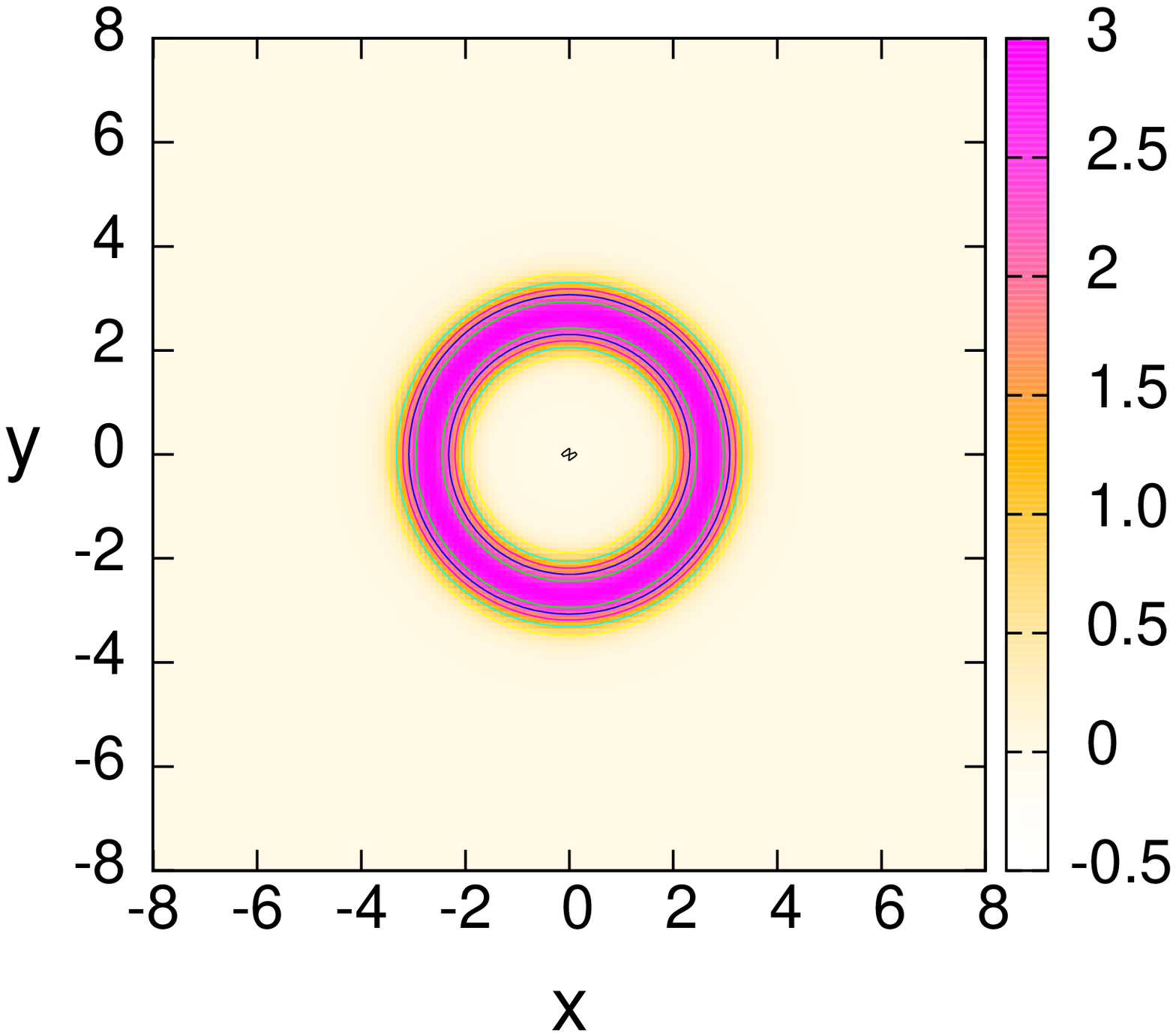}}
{\includegraphics[scale=0.31,angle=-90]{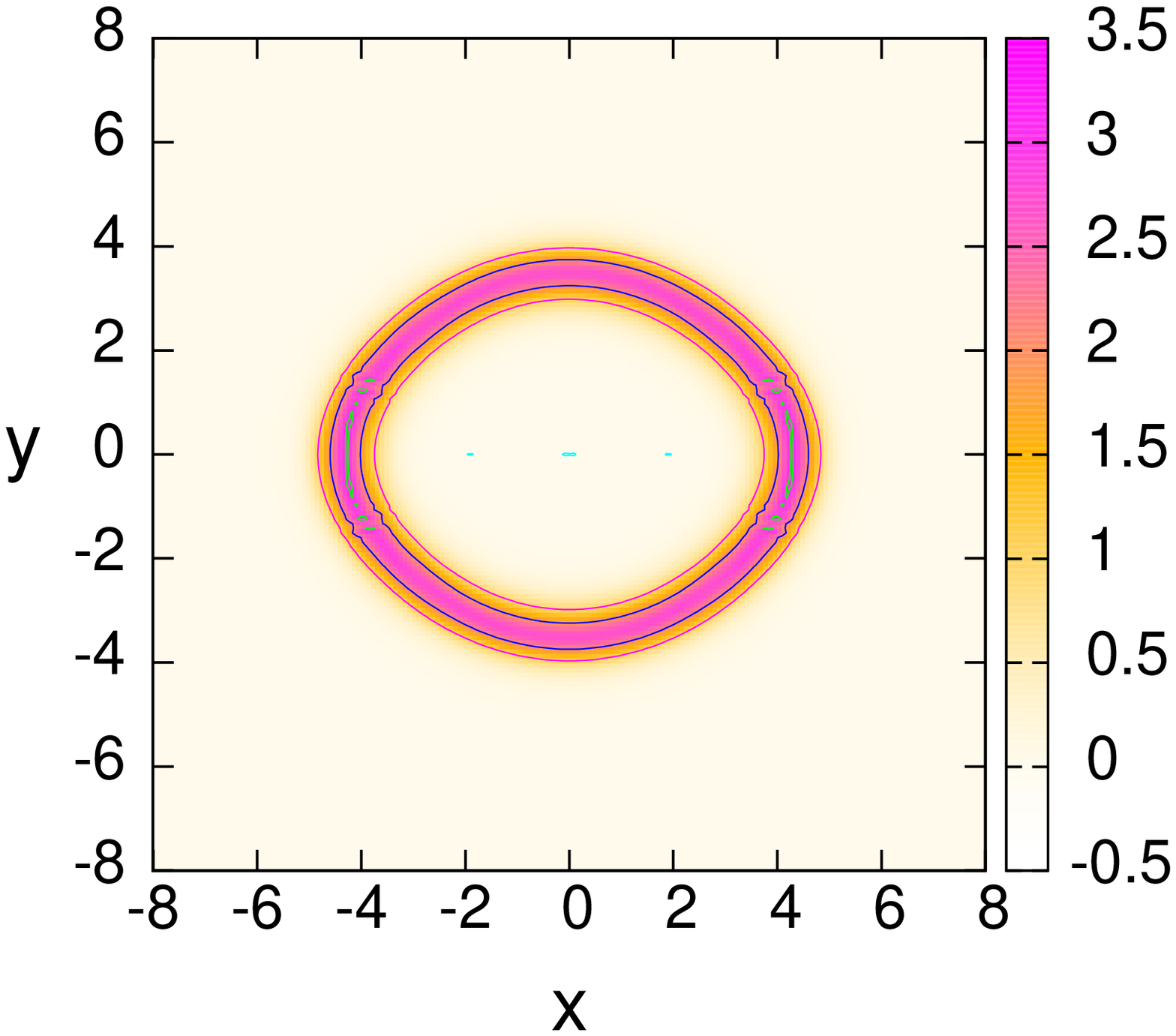}}
\end{center}
\caption{The contour plots of the energy density of the
soliton solutions of the model with double vacuum potential
\re{double} with charges $B=4$ and $B=5$ for $\mu^2=4$ at $\omega =0$ (left panels) and
$\omega=\sqrt 2$ (right panels).}
\end{figure}

\section{Conclusions}
We have studied isospinning soliton solutions of the low-dimensional baby Skyrme model of degree $1\le B \le 5$ with
two types of potential, the "old" model and the "new" double vacuum model. Similar to the case of the isospinning solitons of the
Faddeev--Skyrme model \cite{JHSS,BattyMareike,Acus:2012st}, we used
reformulation of the minimization problem considering the stationary points of the pseudoenergy functional
$F_\omega(\omega)$ which we found numerically without imposing any assumptions about the spatial symmetries.
Our results confirm that
the solitons persist for all range of values of $\omega \le~{\rm min} \{\sqrt 2,\mu\}$, where $\mu$ is the
mass of the scalar excitations, and their qualitative shape is independent of the frequency $\omega$.
Thus, there are two types of instabilities of the solitons, one is due to radiation of the scalar field and another one
is related with destabilization of the rotating solitons by nonlinear velocity terms \cite{JHSS}.

Apart from the above feature, we have found that the critical behavior of isospinning
configurations strongly depends on the structure of the potential term of the model.
The multi-solitons of the "new" baby Skyrme model
remain stable with respect to decay into constituents of smaller charge up to the critical
value of the angular parameter $\omega$ at which the configuration ceases to exist, however
the charge $B>3$ solitons of the "old" model decay into constituents at some value of $\omega$
at which the pseudoenergy of the isospinning configuration per unit charge becomes higher than
the pseudoenergy of the system of isospinning baby Skyrmions of lower charge. Also note that
our results indicate that the solitons of the holomorphic model do not admit isorotations.

Similar results about the critical behavior of the isospinning solitons in the old baby Skyrme model were
reported in a very recent paper \cite{BattyMareike2}.
More systematic investigation of
the spinning baby Skyrmions for various potentials and for larger values of $B$ will be presented elsewhere in our work in
collaboration with  R.~Battye and M.~Haberichter \cite{BHHS}.

\section*{Acknowledgments}
We thank Richard Battye, Mareike Haberichter, Itay Hen, Marek Karliner, Olga Kichakova, Jutta Kunz, Paul Sutcliffe and Wojtek Zakrzewski
for useful discussions and valuable comments. This work was financially supported by Alexander von Humboldt Foundation
in the framework of the Institutes linkage
Programm. Some of the work of Ya.S. was carried out at University of Vilnius
supported by the EU BMU-MID visiting grant. Ya.S. is very grateful to the Institute of Physics, Carl von Ossietzky University
for hospitality in Oldenburg.

%\newpage

\end{document}